\documentclass[manuscript]{acmart}

\AtBeginDocument{%
  }

\setcopyright{acmlicensed}
\copyrightyear{2018}
\acmYear{2018}
\acmDOI{XXXXXXX.XXXXXXX}

\acmJournal{JACM}
\acmVolume{37}
\acmNumber{4}
\acmArticle{111}
\acmMonth{8}

\usepackage{graphicx}   
\usepackage{subcaption} 

\usepackage{enumitem}
\usepackage[normalem]{ulem}
\usepackage{soul}
\usepackage{booktabs}
\usepackage{multirow}
\usepackage{longtable}
\usepackage{pgfplots}
\pgfplotsset{compat=1.18}

\newcommand{\added}[1]{\textcolor{black}{#1}}
\newcommand{\removed}[1]{}

\begin{document}

\title{How Good is \textsc{ChatGPT} in Giving Advice on Your Visualization Design?}


\author{Nam Wook Kim}
\email{nam.wook.kim@bc.edu}
\orcid{0000-0003-4899-6671}
\affiliation{%
  \institution{Boston College}
  \streetaddress{140 Commonwealth Ave}
  \city{Chestnut Hill}
  \state{Massachusetts}
  \country{USA}
  \postcode{02467}
}

\author{Yongsu Ahn}
\email{yongsu.an@bc.edu}
\orcid{0000-0002-5797-5445}
\affiliation{%
  \institution{Boston College}
  \streetaddress{140 Commonwealth Ave}
  \city{Chestnut Hill}
  \state{Massachusetts}
  \country{USA}
  \postcode{02467}
}

\author{Grace Myers}
\email{grace.myers@bc.edu}

\affiliation{%
  \institution{Boston College}
  \streetaddress{140 Commonwealth Ave}
  \city{Chestnut Hill}
  \state{Massachusetts}
  \country{USA}
  \postcode{02467}
}

\author{Benjamin Bach}
\email{benjamin.bach@inria.fr}
\affiliation{%
\institution{INRIA Bordeaux}
  \streetaddress{200 Av. de la Vieille Tour, 33405 Talence}
  \city{Bordeaux}
  \country{France}}

\renewcommand{\shortauthors}{Kim et al.}

\begin{abstract}

Data visualization creators often lack formal training, resulting in a knowledge gap in design practice. Large-language models such as \textsc{ChatGPT}, with their vast internet-scale training data, offer transformative potential to address this gap. \added{In this study, we used both qualitative and quantitative methods to investigate how well ChatGPT can address visualization design questions.} First, we quantitatively compared the \textsc{ChatGPT}-generated responses with anonymous online \textsc{Human} replies to data visualization questions on the VisGuides user forum. Next, we conducted a qualitative user study examining the reactions and attitudes of practitioners toward \textsc{ChatGPT} as a visualization design assistant. Participants were asked to bring their visualizations and design questions and received feedback from both \textsc{Human} experts and \textsc{ChatGPT} in randomized order. Our findings from both studies underscore ChatGPT’s strengths---particularly its ability to rapidly generate diverse design options---while also highlighting areas for improvement, such as nuanced contextual understanding and fluid interaction dynamics beyond the chat interface. Drawing on these insights, we discuss design considerations for future LLM-based design feedback systems.

\end{abstract}



\received{20 February 2007}
\received[revised]{12 March 2009}
\received[accepted]{5 June 2009}

\maketitle

\section{Introduction}

Visualizations are ubiquitous and widely employed by practitioners across various disciplines. However, many of those creating visualizations commonly lack formal training in visualization design and instead acquire the necessary skills on the go, leaning from examples from the internet, online blogs, and other publicly available resources~\cite{esteves2022learned,dvsurveys}. These people may find themselves in situations where they must navigate intricate design choices~\cite{choi2023vislab} and consequently revert to their instincts, drawing from the experiences and observations they have accumulated along the way~\cite{choi2023vislab,bako2022understanding}. Alternatively, they seek feedback from colleagues who can help provide novel perspectives, validate design choices, and challenge assumptions~\cite{choi2023vislab,luther2015structuring}. Nevertheless, not all practitioners have the privilege of accessing such feedback from experienced colleagues.

This paper delves into ways to address this knowledge gap in practical settings. We examine the capabilities of large language models (LLMs) equipped with extensive internet-scale training data, which holds transformative potential~\cite{okerlund2022s}. LLM-based chatbots, exemplified by \textsc{ChatGPT}~\cite{OpenAIChatGPT}, can offer tailored suggestions by analyzing a practitioner's design choices and considering established best practices. This emerging advancement can assist casual and novice practitioners, who may have limited knowledge and resources in data visualization, by providing human-like guidance throughout their design journey. 

However, open questions remain regarding the extent of LLMs' current knowledge of visualization best practices, the quality and practical applicability of their suggestions, and how these suggestions differ from those provided by human counterparts. Addressing these open questions, we have formulated two following research questions to guide our investigation:

\begin{itemize}
    \item \textbf{RQ1 (Quantitative evaluation):} \added{How well does ChatGPT respond to visualization design questions posed in online communities compared to practitioners in the wild?}

    \item \textbf{RQ2 (Qualitative experience):} \added{How do visualization practitioners perceive ChatGPT’s design feedback compared to those offered by seasoned professionals?
    }

\end{itemize}

\begin{figure*}
  \includegraphics[width=\textwidth]{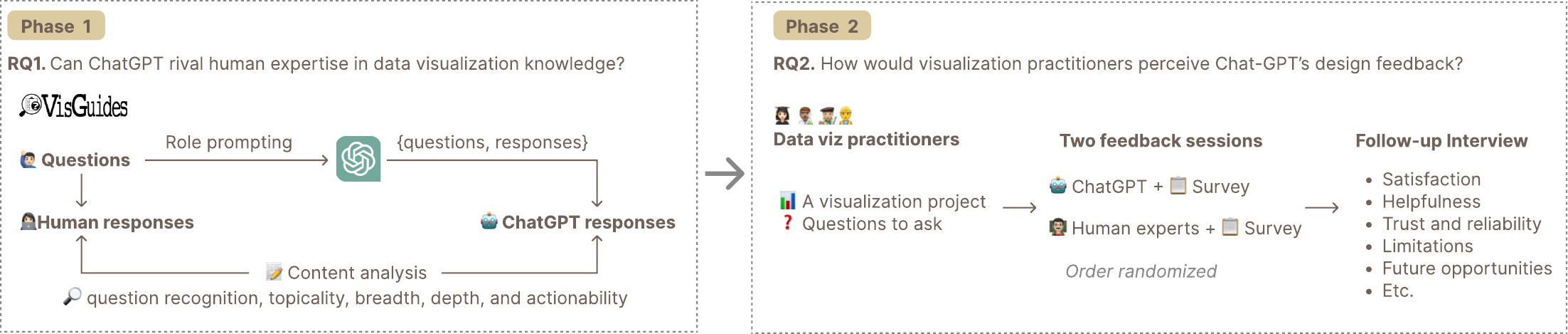}
  \caption{Methodology Overview: The methodology comprises two key phases. In the first phase, questions answered within a forum space by \textsc{Human} respondents are explored and then presented to \textsc{ChatGPT}. The second phase involves a feedback session in which users solicit visualization design feedback from both \textsc{ChatGPT} and \textsc{Human} experts.
}
  \label{fig:methodology-overview}
\end{figure*}

We employ a mixed-methods approach (Fig. \ref{fig:methodology-overview}) to tackle these questions, capturing both breadth (RQ1) and depth (RQ2). \added{RQ1 focuses on a more systematic evaluation using a broader and more realistic knowledge baseline, while RQ2 centers on the qualitative experience of receiving feedback from ChatGPT, assessed against the high standards of domain experts. In our comparisons, seasoned professionals represent trusted colleagues who bring domain expertise and rigor, whereas online practitioners provide a practical, real-world perspective, recognizing that not everyone has access to expert advice~\cite{choi2023vislab}. }
We selected \textsc{ChatGPT} as a representative proxy for LLMs due to its widespread use and generally superior performance compared to other LLMs~\cite{borji2023battle,ahmed2023chatgpt,giannakopoulos2023evaluation,lim2023benchmarking}.

\added{For the first question}, we investigated posts and discussions at the VisGuides forum~\cite{diehl2018visguides}, which captures public questions and responses about visualization design. 
We fed the questions to \textsc{ChatGPT} and compared its responses to the \textsc{Human} responses on VisGuides across six key metrics: coverage, topicality, breadth, clarity, depth, and actionability. We found that ChatGPT's response is comparable to, and often better than, \textsc{Human} responses, especially in providing clear and comprehensive answers that cover all questions asked. In contrast, human responses typically offered specific suggestions, like citing references beyond basic improvements, but showed greater variability due to multiple respondents with diverse backgrounds involved. \added{Given our analysis pertains to specific versions of ChatGPT, we discuss how we can devise an evaluation pipeline that uses LLMs for automatic response scoring.}

\added{For the second question}, we conducted a qualitative user study in which we invited practitioners to participate in a comparative evaluation of \textsc{Human} expert feedback and \textsc{ChatGPT}-generated feedback (GPT-3.5 Turbo). 
For this study, participants were invited to bring their own visualization designs and projects to our sessions for questions and feedback. 
Most study participants preferred discussing with \textsc{Human} experts, valuing their ability to engage in fluid conversations and provide tailored recommendations. However, participants also appreciated the broad knowledge of \textsc{ChatGPT}, particularly its ability for quick brainstorming and ideation from a neutral standpoint, while acknowledging its limitations in deeper contextual understanding and its lack of critical feedback.


Both studies consistently shed light on the strengths and weaknesses of \textsc{ChatGPT} and \textsc{Human}s. \textsc{ChatGPT} excelled at providing broad suggestions, whereas \textsc{Human}s demonstrated a better grasp of nuanced details for more in-depth ideas. A stronger preference for \textsc{Human}s in the feedback study was also linked to the more consistent opinions and fluid interaction dynamics of \textsc{Human} experts. These findings also point to areas for future enhancement of LLMs, such as \added{supporting advanced prompting with better visual perception and fostering proactive and critical discussion capabilities to more effectively handle the intricate contexts of design challenges.}
We distill these lessons into design considerations for future design feedback systems using LLMs. Furthermore, we discuss the potential role of an LLM-based design knowledge companion in visualization education.

Our work represents a meaningful step forward for the field amidst the rise of LLM-driven visualization innovations. The contributions of this work lies in the following:
\begin{itemize}
    \item Evaluation framework for assessing the capabilities of LLMs in providing data visualization design assistance, incorporating both qualitative and quantitative metrics.
    \item  Key insights into the role of LLM-based chatbots, exemplified by \textsc{ChatGPT}, in visualization practice, highlighting their strengths and limitations compared to \textsc{Human} counterparts.
    \item Design considerations for improving the user interface of LLM-based chatbots as collaborative design companions, fostering more effective human-AI interaction. 
\end{itemize}


\section{Related Work}

\subsection{Data Visualization Design Practice}
Data visualization has gone mainstream, frequently used to explore and communicate data in a wide variety of industry sectors. As a result, data visualization practitioners have diverse professions, such as managers, scientists, journalists, and analysts, not necessarily having a specific job title as ``visualization designer'' or ``visualization developer.''~\cite{esteves2022learned} These professionals do not typically have formal training or education in data visualization, while mostly learning necessary visualization skills on the go~\cite{esteves2022learned,dvsurveys}. While decades of empirical research in data visualization have produced fruitful knowledge in effective visualization design, it is unclear how well it is accessible to practice~\cite{choi2021toward,diehl2018visguides}. In fact, it has been reported that many practitioners are not aware of theories and principles from research~\cite{parsons2021understanding}.

As with any creative design profession, visualization practitioners run into challenging situations where they must make decisions among conflicting design alternatives~\cite{choi2023vislab}. Such decisions might include selecting chart types, making competing decisions between aesthetics and functions, picking appropriate color scales, and addressing conflicts with user and business stakeholder needs~\cite{choi2023vislab}. Processes by practitioners to address such design issues are more nuanced and multi-faceted than theoretical process models proposed in the research field~\cite{parsons2021understanding}. They depend on a kind of situated planning that often relies on forms of experimentation and responding to what is happening in the moment~\cite{parsons2021understanding,parsons2020design}. They balance readability and engagement and also consider contextual factors such as whether their visualizations are business-oriented dashboards or public-facing graphics~\cite{parsons2020data}. 

Practitioners frequently resort to their intuition or gut instinct while also looking for examples for inspiration~\cite{choi2023vislab,parsons2021understanding}. Moreover, they also seek feedback from their colleagues to improve their data visualizations~\cite{choi2023vislab,esteves2022learned}.  Such feedback is essential to assess a design and generate revision ideas in the design process, although the fear of criticism and non-anonymity can make people uncomfortable receiving feedback from colleagues. Moreover, self-employed people might struggle to find such colleagues who can provide valuable feedback~\cite{chalkidis2019neural}

Previous research has utilized online crowdsourcing to streamline the feedback collection process, providing rapid access to large, diverse participant pools and enabling faster feedback cycles at reduced costs~\cite{buhrmester2016amazon,haug2021crowd,yen2016social,luther2015structuring}. To address the challenge of insufficient expertise among crowd workers, researchers have explored various techniques to enhance feedback quality, including micro-tasking~\cite{xu2014voyant,luther2014crowdcrit}, implementing predefined rubrics~\cite{yuan2016almost}, providing illustrative examples~\cite{kang2018paragon}, and establishing feedback guidelines~\cite{krause2017critique}. While these efforts have primarily focused on feedback for graphic design projects, recent studies in data visualization have investigated automated methods for providing perceptual feedback through image processing techniques~\cite{shin2023perceptual} and developing structured methods for collecting empirically-driven quantitative feedback~\cite{choi2023vislab}.

\textit{In this work, we examine how an LLM-based chatbot, trained with extensive knowledge in the wild, can serve as a design companion to provide knowledge and feedback to varying backgrounds of practitioners. Can LLMs replace the feedback gathering process? We aim to bridge the gap between LLMs' theoretical potential and their practical application in data visualization, a critical endeavor in the evolving landscape of artificial intelligence within the field.
}

\subsection{Evaluating LLMs' Knowledge Capacity}
LLMs, such as BERT~\cite{devlin2018bert}, GPT~\cite{brown2020language,openai2023gpt}, and Llama~\cite{touvron2023llama}, have rapidly gained popularity and adoption across both industry and academia~\cite{bommasani2021opportunities,zhao2023survey}. Leveraging massive amounts of text data and computational power, these foundation models have achieved impressive performance on a wide variety of tasks, from generating human-like text to high-quality images~\cite{radford2019language,radford2018improving,wei2022emergent}.  However, the scale and complexity of these models make it difficult to fully understand the scope and limitations of their language capabilities~\cite{bender2021dangers}.

A number of benchmark tasks have been proposed to assess the general knowledge and reasoning capabilities of language models in a more holistic manner~\cite{liang2022holistic,chang2023survey}.  These include tasks that require models to answer broad-coverage factoid questions based on pre-existing knowledge~\cite{liang2022holistic}. Other benchmarks evaluate numeric, analogical, and logical reasoning skills across different modalities~\cite{qin2023chatgpt,zhang2023multimodal}, while some others examine the ability of LLMs to generalize knowledge to new concepts~\cite{wei2022emergent}. Assessing models on such diverse reasoning skills can provide deeper insight into their adaptability, knowledge gaps, and limitations. In addition to assessing such capabilities, an important area of research has focused on evaluating potential harms and biases in large language models. Some studies have performed targeted adversarial evaluations to expose problematic biases encoded in the models' training data~\cite{nadeem2021stereoset,tan2019assessing,gehman2020realtoxicityprompts}.

In addition to generic capabilities, research has been carried out to investigate LLMs' discipline-specific knowledge and skills. The inherent intricacies and nuances of specific domains necessitate tailored evaluation methods. For instance, researchers explored students' and teachers' usage of and perceptions toward LLMs, investigating opportunities and potential benefits for education~\cite{kasneci2023chatgpt,baidoo2023education,tlili2023if}. Researchers also tested LLMs' ability to pass exams in specific domains like medicine~\cite{gilson2023does}, law~\cite{fei2023lawbench,choi2023chatgpt}, and business~\cite{fortune2023chatgpt}. Others have evaluated LLMs' ability to carry out real-world tasks. Examples include analyzing diagnostic abilities by answering questions about medications, symptoms and treatments~\cite{lee2023benefits}; producing research hypotheses~\cite{lahat2023evaluating}; predicting legal judgment and recommending case citations~\cite{chalkidis2019neural}; and predicting chemical properties~\cite{pan2023large}.

\added{Many recent studies have explored the applicability of LLMs in the realm of data visualization. Inala et al. examine the potential impact of generative AI on the data analysis process, encompassing data collection, transformation, and exploration~\cite{inala2024data}. DataFormulator 1 \& 2 focus on facilitating data transformation workflows for creating data visualizations~\cite{wang2023data,wang2024data}. On the other hand, several tools leverage LLMs to provide tailored natural language interfaces for generating data visualizations, including LIDA~\cite{dibia2023lida}, DynaVis~\cite{vaithilingam2024dynavis}, and ChartGPT~\cite{tian2024chartgpt}. 
Other studies have evaluated LLMs' capabilities in eliciting design preferences~\cite{wang2024dracogpt}, assisting in chart interpretation~\cite{wang2024howaligned}, detecting misleading charts~\cite{lo2024good,alexander2024can}, performing visualization literacy tasks~\cite{bendeck2024empirical,choe2024enhancing}, generating visualization code~\cite{chen2024viseval}, and troubleshooting visualization generation code~\cite{shen2024ask}. Furthermore, user studies have examined how ChatGPT is utilized in classrooms~\cite{kim2024chatgpt}, visualization practitioners' perceptions of text-to-image models~\cite{di2023doom}, the impact of AI assistance on analysis workflows~\cite{gu2024data}, and its effects on visualization coding performance~\cite{vaithilingam2022expectation}.}



\added{\textit{The two pillars of data visualization skills are design and implementation. Most previous studies have focused on data visualization generation and implementation. In contrast, this work focuses on evaluating the design feedback capabilities of LLMs. Similar efforts have recently been explored in user interface design, where researchers have leveraged existing guidelines~\cite{duan2024generating} and critique datasets~\cite{duan2024uicrit} to perform design evaluations. In our study, we use real-world design questions from practitioners to conduct evaluations.}}


\section{Assessing \textsc{ChatGPT}'s Competence in Data Visualization Knowledge}

As a step toward assessing \textsc{ChatGPT}'s data visualization knowledge, we were interested in its ability to respond to real-world data visualization questions in comparison to \textsc{Human}s.

\subsection{Data Collection}

We used VisGuides~\cite{diehl2018visguides} as our main source for visualization questions. VisGuides is a discussion forum focused on visualization guidelines, where practitioners from various backgrounds, such as scientists, designers, engineers, and students, ask and respond to questions regarding their visualizations and fundamental visualization principles.

We identified a total of 226 questions within the VisGuides repository. Among these, we selected \textbf{119 questions} for the study based on specific inclusion criteria. These criteria included the presence of a question in the post, the requirement for a visualization description to possess sufficient visual encoding (due to \textsc{ChatGPT}'s inability to process images at the time of our analysis), the need for questions to be comprehensible and well-structured (excluding overly generic inquiries like 'What do you think?' or 'Is this good enough?'), and the necessity of having at least one \textsc{Human} response. Two of the authors jointly reviewed and unanimously approved the inclusion of these 119 questions.

We discovered that Visguides primarily featured two types of questions, broadly divided into two categories: \textit{design feedback questions} (86) and \textit{visualization guideline questions} (33). The former aimed to enhance users' personal visualizations (see Figure \ref{fig:visguide-questions}B \& C), while the latter sought to comprehend visualization guidelines or principles (see Figure \ref{fig:visguide-questions}A). For a more comprehensive taxonomy of questions on the platform, Diehl et al. provide detailed insights~\cite{diehl2021visguided}.

After compiling the list of qualifying questions, we fed each query to \textbf{\textsc{ChatGPT} 3.5 in May 2023} and with images to \added{\textbf{ChatGPT 4 with Vision in May 2024} (accompanied by images)}, accompanied by a role-playing prompt designed as follows:

\begin{normalsize}
\begin{verbatim}
Please act as if you are a visualization expert.

Can you respond to this question delimited by ///.

 ///
 
 {Question}
 
 ///

 [...]

Please format your reply as JSON objects with keys: Question and Response.
  {
    Question: xxx
    Response: xxx
  }
There may be more than one question, format each as a new object.
\end{verbatim}
\end{normalsize}

To streamline our analysis, we created a spreadsheet where each row included a question, the corresponding \textsc{ChatGPT} responses (\added{3.5 and 4V}) and a compilation of \textsc{Human} responses.
Each question had 1.54 \textsc{Human} responses on average ($Std.dev$ = 1.03).
The average approximate word count of the \textsc{Human} responses was 235.35 ($Std.dev$ = 289.62) with a maximum of 1721 words and minimum of 13 words. The average approximate word count of the \textsc{ChatGPT} 3.5 responses was 463.42 ($Std.dev$ = 189.34) with a maximum of 1192 words and minimum of 115 words. \added{For \textsc{ChatGPT} 4, the responses averaged around 255.03 words ($Std.dev$ = 88.58), with lengths spanning from 110 to 491 words.}

\subsection{Analysis Method}
Two researchers performed an open coding process on a randomly selected 20\% sample of the questions and the corresponding \textsc{ChatGPT} and \textsc{Human} responses. \added{They iteratively agreed on and refined six evaluation metrics: coverage, topicality, breadth, depth, clarity, and actionability. Since design is a complex problem-solving task, we believe that qualitative human-centered metrics are more appropriate for capturing the holistic quality of responses compared to traditional utility metrics such as accuracy. These metrics were assessed using a Likert scale that ranged from 1: Very poor to 5: Very good. Table \ref{tab:rubrics} presents our codebook of metric definitions and scoring rubrics.}

\begin{table}[ht]
\small
\centering
\caption{\added{Metrics: Definitions and Scoring Rubrics} \label{tab:rubrics}}
\label{tab:rubrics}
\begin{tabular}{@{}p{0.35\textwidth} p{0.6\textwidth}@{}}
\toprule
\textbf{Metric \& Definition} & \textbf{Score \& Description} \\
\midrule
\multirow{5}{=}{
  \textbf{Coverage} \textit{assesses whether the response addresses all questions posed by the user.}
}
& \textbf{1:} Addresses 0--20\% of the user’s questions. \\
& \textbf{2:} Addresses 20--40\% of the user’s questions. \\
& \textbf{3:} Addresses 40--60\% of the user’s questions. \\
& \textbf{4:} Addresses 60--80\% of the user’s questions. \\
& \textbf{5:} Addresses 80--100\% of the user’s questions. \\
\midrule
\multirow{5}{=}{
  \textbf{Topicality} \textit{indicates how closely the response stays on topic.}
}
& \textbf{1:} Mostly or entirely off-topic. \\
& \textbf{2:} Partially addresses the topic but frequently digresses. \\
& \textbf{3:} Somewhat relevant; minor deviations from the topic. \\
& \textbf{4:} Mostly on-topic; occasional tangents. \\
& \textbf{5:} Fully on-topic with no noticeable digressions. \\
\midrule
\multirow{5}{=}{
  \textbf{Breadth}\textit{measures how widely a response explores various ideas, concepts, options, or perspectives.}
}
& \textbf{1:} Very narrow; offering no alternative ideas. \\
& \textbf{2:} Covers a small range of ideas, omitting key concepts. \\
& \textbf{3:} Moderately diverse but limited in scope.\\
& \textbf{4:} Addresses multiple angles with reasonable detail. \\
& \textbf{5:} Very wide-ranging; covers a broad spectrum of ideas thoroughly. \\
\midrule
\multirow{5}{=}{
  \textbf{Depth} \textit{evaluates the extent of explanations, expertise, and valuable insights in a response.}
}
& \textbf{1:} Very minimal detail; lacks substance or justification. \\
& \textbf{2:} Superficial treatment with limited insight. \\
& \textbf{3:} Moderately detailed; some analysis or explanation. \\
& \textbf{4:} Good explanation with solid reasoning and insight. \\
& \textbf{5:} Deeper explanation, demonstrated with examples or expertise \\
\midrule
\multirow{5}{=}{
  \textbf{Clarity} \textit{assesses how easily a reader can understand the response, considering conciseness, organization, and readability.}
}
& \textbf{1:} Very unclear; difficult to follow or disorganized. \\
& \textbf{2:} Somewhat unclear; main ideas partially obscured. \\
& \textbf{3:} Moderately clear; understandable with some effort. \\
& \textbf{4:} Clear, well-structured, and easy to read. \\
& \textbf{5:} Flows seamlessly with minimal effort to comprehend. \\
\midrule
\multirow{6}{=}{
  \textbf{Actionability} \textit{assesses whether the response offers guidance readily implementable in a visualization (N/A for guideline-only questions).}
}
& \textbf{1:} No actionable guidance or next steps. \\
& \textbf{2:} Little actionable guidance; mostly explanations. \\
& \textbf{3:} Some actionable guidance; partially implementable. \\
& \textbf{4:} Clear, actionable guidance that can be followed easily. \\
& \textbf{5:} Highly actionable; provides explicit step-by-step instructions. \\

\bottomrule
\end{tabular}
\end{table}

The researchers initially assigned scores to the 20\% sample, engaging in discussions to resolve any discrepancies and establish consensus definitions and criteria for each score within every metric to ensure consistency. Following this, one of the researchers proceeded to score the remaining question responses. Throughout this coding process, they held continuous meetings to discuss any new issues and questions to ensure consistency and quality of codes and scores, refining metric definitions and score applicability.

Given that some questions elicited multiple \textsc{Human} responses while others received only one, we made them to be evaluated with a single score to assess the collective quality of the \textsc{Human} responses. \added{This score captures the cumulative strengths of all sub-responses, recognizing that a single high-quality sub-response can adequately address the questioner’s needs even if others are less informative. We performed Mann–Whitney U tests to compare \textsc{Human} composite scores with \textsc{ChatGPT} response quality scores.}

Furthermore, any elements of the \textsc{Human} responses that extended beyond the capabilities of \textsc{ChatGPT}, such as providing citations to related literature, links to external resources, visual examples, or posing follow-up questions, were recorded. We also categorized whether each question sought feedback on the questioner's visualizations or inquired about general visualization design knowledge.

\subsection{Analysis Decisions}
\paragraph{Why Did We Use VisGuides?}
We explored other sources of data visualization questions in the wild, including StackOverflow, the \texttt{\#help-general} channel in the Slack workspace of The Data Visualization Society~\cite{datavissociety}, and \texttt{r/dataisbeautiful} on Reddit. StackOverflow mainly focuses coding-related questions, which differ from our focus on design knowledge questions. The Data Visualization Society's Slack channel, limited by a lack of a paid license, does not archive questions older than 90 days, and the queries are mostly technical, such as using specific software tools to create charts or visual effects. Reddit's \texttt{r/dataisbeautiful} channel aligns more closely with our objectives, but a thorough review revealed that it contains more discussions and occasional critiques of web-found charts than actual questions. Hence, we decided to focus on the questions in the VisGuides forum.

\paragraph{Why Did We Choose \textsc{ChatGPT} Over Other Chatbots?}

We selected \textsc{ChatGPT} as a representative of LLM-based chatbots for several reasons. While these chatbots use the same transformer architectures, ChatGPT is the most widely adopted in practice. Also, in our early pilot experimentation with Bard (now Gemini), Claude, and Bing (which uses the same GPT model as \textsc{ChatGPT}), \textsc{ChatGPT} generally provided responses that were better aligned in terms of style and content. For example, Bing often veered off by drawing information from various web sources, which were frequently inferior to \textsc{ChatGPT}'s internal knowledge responses. 
Furthermore, Bard's recommendations were narrower in scope and often not applicable, like suggesting the addition of a trend line to a Gantt Chart. Overall, our impression was that these systems are mostly comparable, with ChatGPT having a slight edge over the others. Several studies also indicated that \textsc{ChatGPT} demonstrates superior performance compared to other LLMs in terms of a variety of task questions involving reasoning, facts, and fairness~\cite{borji2023battle,ahmed2023chatgpt}.
\added{That said, we discuss an automatic evaluation pipeline, outlined in the discussion section, to enable comparisons of different models without requiring human intervention.}


\paragraph{Why Did We Not Use Different Prompting Methods?}

\textsc{ChatGPT} and other LLM-based chatbots are designed to follow human instructions~\cite{wu2023brief,liu2021pre}. Different prompting techniques, such as chain-of-thought~\cite{wei2022chain} and self-consistency~\cite{wang2022self}, can lead to varied responses. To ensure a more objective evaluation, we used neutral prompts without any specific instructions (e.g., making the response concise) other than the role prompting. Moreover, we assume that users typically would not employ complex prompting strategies and would most likely use zero-shot queries. Thus, we merely forwarded the question from VisGuides to \textsc{ChatGPT}.

\paragraph{Why use The Six Quality Metrics?}
Our goal was to assess the semantic quality of responses to visualization design questions---how well they conveyed meaningful, useful, and understandable knowledge—via human evaluation. While quantifiable measures (e.g., sentence structure, rhetorical style) can provide quick and repeatable approximations of text quality \cite{van2021human}, research has shown that they do not fully capture the complex and composite nature of constructs like diversity by quantifiable metrics \cite{van2018measuring} and do not often correlate with how humans assess text quality \cite{scott2006nlg, reiter2018structured}. Based on a review of literature on the human evaluation of automatically generated texts, we selected metrics that cover distinct aspects of intrinsic quality. Coverage and topicality capture the nuanced aspects of relevance—how the response answers the user’s questions directly \cite{van2021human, sreedhar2024canttalkaboutthis, awasthi2023humanely}; breadth and depth reflect the richness of ideas and reasoning \cite{wang2023mathcoder}; actionability assesses practical applicability and informativeness \cite{van2021human}; and clarity is a widely used measure to assess readability, organization, and conciseness \cite{jiang2024large, awasthi2023humanely}. Overall, this framework offers a meaningful foundation for evaluating LLM performance in visualization design, with future work potentially extending it to include response tone, prioritization, or interaction quality.

\subsection{Results}

\begin{figure}
  \includegraphics[width=\linewidth]{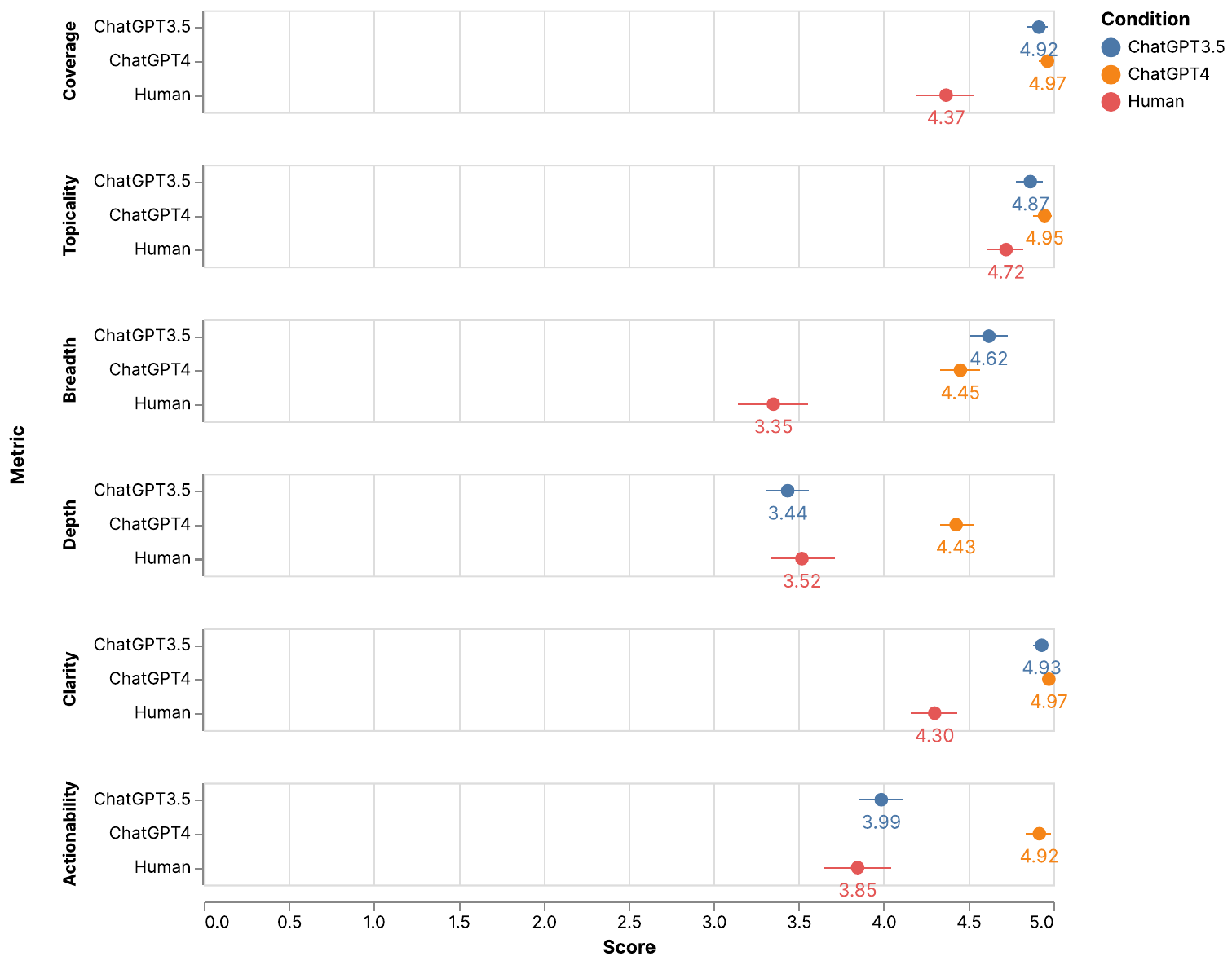}
  \caption{\added{Mean scores (dots) and confidence intervals (error bars) across six metrics—Actionability, Breadth, Clarity, Coverage, Depth, and Topicality—for responses from \textsc{ChatGPT}~3.5, \textsc{ChatGPT}~4, and \textsc{Human}. Overall, \textsc{ChatGPT}~4 shows slightly higher mean scores than \textsc{ChatGPT}~3.5 and \textsc{Human}. \textsc{Human} responses exhibit higher variance but often perform comparably to \textsc{ChatGPT}. } }

  \label{fig:score-result}
\end{figure}

\begin{figure*}
  \includegraphics[width=.95\textwidth]{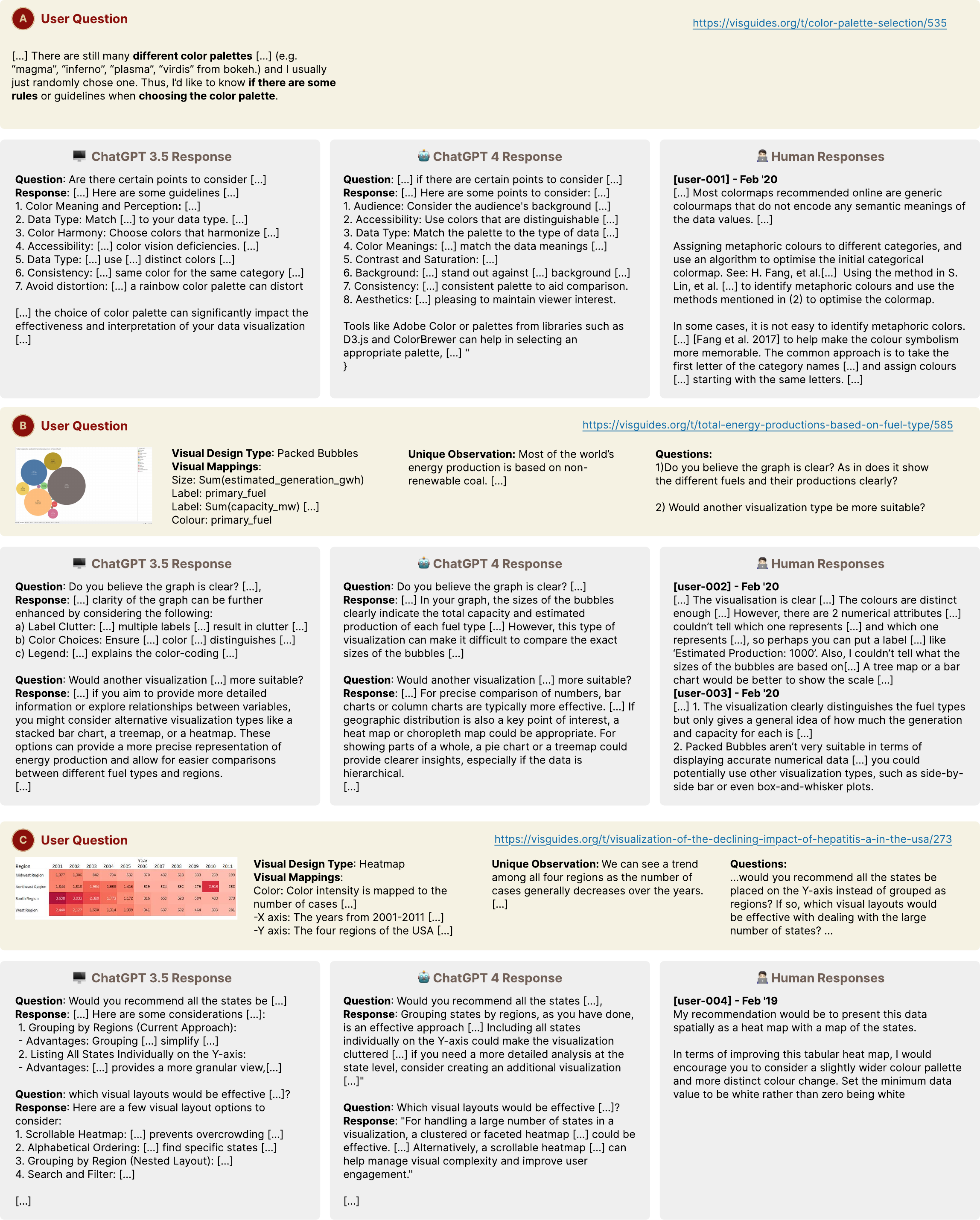}
  \vspace{-1em}
  \caption{Three selected and edited examples of questions and responses from the VisGuides forum. The \textsc{ChatGPT} responses consistently demonstrate breadth, topicality, and coverage. \textsc{ChatGPT}~4, in particular, showcases greater expertise by offering more in-depth and actionable advice in an authoritative manner, rather than merely listing ideas. The \textsc{Human} response ratings are more varied, as evident when comparing the first and last examples. Full questions and responses are provided in the supplement. Respondent IDs have been anonymized.}
 
  \label{fig:visguide-questions}
\end{figure*}

\subsubsection{Ratings on Quality Metrics}

Figure \ref{fig:score-result} shows the scores of response quality metrics for both \textsc{ChatGPT} and \textsc{Human} responses. We describe the findings for each quality metric as below.

\paragraph{Coverage}

\added{\textsc{ChatGPT}~4 achieved the highest score of 4.97 ($std.dev$ = 0.22), closely followed by \textsc{ChatGPT}~3.5 with 4.92 ($std.dev$ = 0.33). \textsc{Humans} scored lower at 4.37 ($std.dev$ = 0.95). Mann–Whitney U tests revealed that the difference between both \textsc{ChatGPT} models and \textsc{Humans} was statistically significant ($p < 0.001$), while the comparison between \textsc{ChatGPT}~3.5 and \textsc{ChatGPT}~4 was not significant ($p = 0.125$).}

The lower score often stemmed from the \textsc{Human} respondent's failure to address all facets of the question. For instance, in Figure \ref{fig:visguide-questions}C, the user raised a two-part question, \textit{``...would you recommend all the states be placed on the Y-axis instead of grouped as regions? If so, which visual layouts would be effective with dealing with the large number of states?...''}. However, the \textsc{Human} response only addressed the second part of the question, suggesting a heat map as a useful visual layout.
 
\paragraph{Topicality} 

\added{\textsc{ChatGPT}~4 achieved the highest score of 4.95 ($std.dev$ = 0.31), closely followed by \textsc{ChatGPT}~3.5 with 4.87 ($std.dev$ = 0.45). \textsc{Humans} scored slightly lower at 4.72 ($std.dev$ = 0.60). Mann–Whitney U tests revealed that \textsc{ChatGPT}~4 significantly outperformed \textsc{Humans} ($p < 0.001$), and \textsc{ChatGPT}~3.5 also showed a statistically significant advantage over \textsc{Humans} ($p = 0.014$). However, the comparison between \textsc{ChatGPT}~3.5 and \textsc{ChatGPT}~4 was not significant ($p = 0.062$).}

While all conditions performed well in addressing the correct question, there were moments when both \textsc{ChatGPT} and \textsc{Human}s strayed from the main topic. For instance, when a user inquired about the application of visual variables in VR visualizations\footnote{\href{https://visguides.org/t/visual-variables-for-visualizations-in-vr/143}{visguides:visual-variables-in-vr}}, \textsc{ChatGPT} \added{3.5} not only addressed this question but also came up with a hypothetical question asking what factors to consider when choosing such visual variables. \textsc{Human} respondents also occasionally veered off-topic by offering unsolicited advice and extending their responses beyond the original question's scope.


\paragraph{Breadth} 
\added{\textsc{ChatGPT}~3.5 achieved the highest score of 4.62 ($std.dev$ = 0.61), slightly outperforming \textsc{ChatGPT}~4, which scored 4.45 ($std.dev$ = 0.66). \textsc{Humans} scored significantly lower at 3.35 ($std.dev$ = 1.13). Mann–Whitney U tests revealed that both \textsc{ChatGPT}~3.5 and \textsc{ChatGPT}~4 significantly outperformed \textsc{Humans} ($p < 0.001$ for both comparisons). Additionally, \textsc{ChatGPT}~3.5 demonstrated a statistically significant advantage over \textsc{ChatGPT}~4 ($p = 0.025$).}

For example, in Figure \ref{fig:visguide-questions}A, when the user inquired about color palette guidelines, \textsc{ChatGPT} provided an extensive list of potential approaches, encompassing data type, color harmony, accessibility, etc. In contrast, the \textsc{Human} respondent to this question concentrated primarily on the creation of domain-specific color maps, offering in-depth insights into the process.


\paragraph{Depth} 
\added{For the depth metric, \textsc{ChatGPT}~4 achieved the highest score of 4.43 ($std.dev$ = 0.56), outperforming both \textsc{ChatGPT}~3.5, which scored 3.44 ($std.dev$ = 0.71), and \textsc{Humans}, who scored 3.52 ($std.dev$ = 1.07). Mann–Whitney U tests revealed that \textsc{ChatGPT}~4 significantly outperformed both \textsc{ChatGPT}~3.5 ($p < 0.001$) and \textsc{Humans} ($p < 0.001$). However, there was no significant difference between \textsc{ChatGPT}~3.5 and \textsc{Humans} ($p = 0.518$).}


\added{There were intriguing and consistent behavioral patterns across the conditions. For example, \textsc{ChatGPT}~3.5 responses tended to provide general, surface-level improvements applicable to any visualization, such as adding labels and legends (Figure \ref{fig:visguide-questions}). While these suggestions were often hit or miss, they were effective when the questions sought generic advice. In contrast, \textsc{ChatGPT}~4 exhibited a more human-like behavior, offering focused suggestions in a more authoritative manner, showcasing expertise rather than merely enumerating a list of ideas, as illustrated in Figure \ref{fig:visguide-questions}B and C.}

\added{Good \textsc{Human} responses were similar to \textsc{ChatGPT}~4, often even more nuanced}, as demonstrated by the recommendation to \textit{``put a label before the numerical attribute like `Estimated Production: 1000'} (Figure \ref{fig:visguide-questions}B).'' This detailed guidance directly pertains to the unique aspects of the visualization in question. However, \textsc{Human} responses also \textit{displayed greater variability}. For instance, when asked about color clarity and alternative data representation to reduce clutter, a \textsc{Human} response simply confirmed the current design choices are fine\footnote{\href{https://visguides.org/t/latitude-distribution-of-solar-and-wind-farms-uk/815}{visguides:latitude-distribution-of-solar-and-wind-farms-uk}}. The lower scores of such poor quality single responses pulled the average down.


\paragraph{Clarity} 

\added{Notably, \textsc{ChatGPT} consistently presented its responses in a well-structured list, which is a clear advantage of machine-generated responses. This contributed to \textsc{ChatGPT}~4 achieving the highest clarity score, averaging 4.97 ($std.dev$ = 0.20), closely followed by \textsc{ChatGPT}~3.5 with an average score of 4.93 ($std.dev$ = 0.28).} \textsc{Human} responses also demonstrated good clarity, with an average score of 4.30 ($std.dev$ = 0.80). However, \textsc{Humans} received lower ratings when the responses were overly brief, making it difficult to discern specific reference points. \added{Mann–Whitney U tests revealed that both \textsc{ChatGPT} models significantly outperformed \textsc{Humans} ($p < 0.001$), while the difference between \textsc{ChatGPT}~4 and \textsc{ChatGPT}~3.5 was not statistically significant ($p = 0.093$).}



\paragraph{Actionability}  
\added{\textsc{ChatGPT}~4 achieved the highest score, averaging 4.92 ($std.dev$ = 0.38), followed by \textsc{ChatGPT}~3.5 with an average score of 3.99 ($std.dev$ = 0.60). \textsc{Human} responses had the lowest average score of 3.85 ($std.dev$ = 0.94). Overall, \textsc{ChatGPT}~4 performed exceptionally well in providing actionable recommendations, as its responses consistently included focused, practical advice. By contrast, \textsc{ChatGPT}~3.5 often provided broader, less specific suggestions, which reduced its scores due to the generic or vague nature of the advice (Figure \ref{fig:visguide-questions}B and C), although the list of recommendations often made sense (Figure \ref{fig:visguide-questions}A).} The actionability scores for \textsc{Human} responses tend to be low when the users refrain from offering recommendations and instead validate all of the user's choices\footnote{\href{https://visguides.org/t/comparison-of-type-a-b-acute-hepatitis-from-tycho-dataset/406}{visguides:comparison-of-acute-hepatitis-from-tycho-dataset}}. \added{Mann–Whitney U tests confirmed that \textsc{ChatGPT}~4 significantly outperformed both \textsc{ChatGPT}~3.5 ($p < 0.001$) and \textsc{Humans} ($p < 0.001$). However, there was no significant difference between \textsc{ChatGPT}~3.5 and \textsc{Human} responses ($p = 0.416$).}




\subsubsection{Further Observations in \textsc{Human} and \textsc{ChatGPT} Responses}

\paragraph{\textsc{Human}s tend to provide external resources}---43 out of 119 (36.13\%). \textsc{Human} responses included elements that \textsc{ChatGPT} would not provide unless explicitly instructed; even when asked, it is known to frequently provide incorrect citations~\cite{frosolini2023reference}. This additional content encompassed references to academic research, links to related articles and websites, citations of studies, and the inclusion of informative video links. For instance, in Figure \ref{fig:visguide-questions}A, asking for guidelines on choosing color palette, the responses include references to three different academic research papers including ``H. Fang, et al.''. In another case, when the question was about using blow-apart effects, the \textsc{Human} respondent embedded an educational video related to the subject\footnote{\href{https://visguides.org/t/the-blow-apart-effect/71}{visguides:the-blow-apart-effect}}. 

\paragraph{Multiple \textsc{Human} responses tend to complement each other}---Most questions had a single response (84), while 35 questions had multiple responses. Average ratings generally favored responses with multiple contributors across all categories, except for topicality. Significant rating differences were observed in coverage, breadth, depth, and actionability. Single-respondent questions averaged 4.25 ($std.dev$ = 1.02) in coverage, 3.06 ($std.dev$ = 1.05) in breadth, 3.35 ($std.dev$ = 1.05) in depth, and 3.72 ($std.dev$ = 0.96) in actionability. Conversely, multi-respondent questions scored higher, averaging 4.66 (\added{*}, $std.dev$ = 0.67) in coverage, 4.06 (\added{**}, $std.dev$ = 0.98) in breadth, 3.94 (\added{**}, $std.dev$ = 0.98) in depth, and 4.26 (\added{*}, $std.dev$ = 0.74) in actionability. When there's only one response to a question, the quality of that response becomes heavily reliant on the individual respondent, leading to significant variability in quality, as seen in Figure \ref{fig:visguide-questions}A (high quality) versus Figure \ref{fig:visguide-questions}C (low quality). However, when multiple respondents contribute, they complement each other and compensate for areas in which one respondent might fall short (Figure \ref{fig:visguide-questions}B).


\paragraph{\textsc{Human} response rating improves for visualization guideline questions}---For \textsc{ChatGPT} responses, the differences in ratings between feedback and visualization guideline questions consistently remained small. \added{For \textsc{ChatGPT}~4, there was a significant difference in the depth metric ($p = 0.014$), with guideline questions receiving higher ratings (Mean = 4.64, $std.dev$ = 0.48) compared to feedback questions (Mean = 4.35, $std.dev$ = 0.57). For \textsc{ChatGPT}~3.5, the only significant difference was observed in the coverage metric ($p = 0.002$), where feedback questions scored higher (Mean = 4.97, $std.dev$ = 0.24) compared to guideline questions (Mean = 4.79, $std.dev$ = 0.48). Across other metrics, no significant differences were detected, highlighting consistent performance regardless of the question type.}

\added{However, we observed more significant differences in \textsc{Human} responses between guideline and design feedback questions. For the breadth metric, guideline questions scored significantly higher (Mean = 4.12, $std.dev$ = 0.95) compared to feedback questions (Mean = 3.06, $std.dev$ = 1.05, $p < 0.001$). Similarly, for the coverage metric, guideline questions achieved a higher score (Mean = 4.76, $std.dev$ = 0.43) than feedback questions (Mean = 4.22, $std.dev$ = 1.05, $p = 0.015$). The depth metric also showed a notable difference, with guideline questions receiving a substantially higher score (Mean = 4.24, $std.dev$ = 0.95) compared to feedback questions (Mean = 3.24, $std.dev$ = 0.98, $p < 0.001$). Overall, \textsc{Human} responses to guideline questions tended to be more comprehensive and detailed. For other metrics, no significant differences were observed.}

Guideline questions frequently require broader perspectives and readily accessible design knowledge, with 84.4\% (27/32) of \textsc{Human} responses citing external references, in contrast to design feedback questions, where only 18.39\% (16/87) of \textsc{Human} responses did so. Conversely, feedback questions entail an understanding of domain-specific data and tasks, potentially making it challenging to offer comprehensive insights. For instance, a question like ``Is this color scheme suitable?'' requires a deep understanding of the domain. Conversely, questions such as the one about gendered colors in visualizations\footnote{\href{https://visguides.org/t/use-of-gendered-colours-in-visualization-a-guideline-or-a-personal-principle/999}{{visguides:use-of-gendered-colours-in-visualization}}} can easily elicit a variety of various viewpoints and existing resources.






\paragraph{Question specificity contributes to response variability}---While we did not quantify this, we observed \textsc{ChatGPT}'s response might be susceptible to how specific and clear the question was. Figure \ref{fig:visguide-questions}B shows such an example when the user query was ``do you believe the graph is clear?''
Other examples of less specific user questions include: ``is my color map optimal?'' and ``how can my visual design be improved?''\footnote{\href{https://visguides.org/t/map-and-bar-visualization/841}{visguides:map-and-bar-visualization}}. Such commonly encountered ill-specified questions can pose a challenge for LLMs, as they may begin suggesting irrelevant feedback instead of asking clarifying questions.
On the other hand, in Figure \ref{fig:visguide-questions}C, the question is more specific, resulting in more meaningful options and insights offered by \textsc{ChatGPT}. 

\textbf{\textsc{ChatGPT}~\added{3.5}'s color vision deficiency was usually not problematic}---Color-related questions were prominent. Out of the 119 questions, 32 were centered on color-based design feedback (e.g., ``Is the choice of the colour scheme appropriate?'' \footnote{\href{https://visguides.org/t/area-chart-generation-capacity-over-the-years-for-each-power-fuel-type/810}{visguides:area-chart-generation-capacity-over-the-years}}). Users rarely mentioned the specific colors used in their visualizations, instead providing only visual encoding information in their questions. Consequently, \added{\textsc{ChatGPT}~3.5 lacked access to exact color information due to its inability to comprehend images.} Despite this limitation, there were no significant differences across all metrics in \textsc{ChatGPT}~3.5's responses.


\subsection{Takeaways}

Our analysis shed light on \textsc{ChatGPT}'s cability to respond to data visualization queries compared to \textsc{Human} counterparts. 
\textsc{ChatGPT} clearly excelled in breadth, exploring a wide range of ideas and concepts, surpassing \textsc{Human} responses. Clarity was another area where \textsc{ChatGPT} scored highly, benefiting from its structured response format. \added{\textsc{ChatGPT}~4 outperformed \textsc{ChatGPT}~3.5 by providing more focused and in-depth responses, aided by its image comprehension capabilities.} Some \textsc{Human} responses resembled those of \textsc{ChatGPT}~4, often enhancing their quality by including reference resources and complementing one another in multi-respondent scenarios.



\begin{table*}[]
\small
\resizebox{\textwidth}{!}{%
\begin{tabular}{lllllll}

\toprule
Participant ID & Role & Years in Data Vis & Gender & Age range & Racial background & Frequency of \textsc{ChatGPT} usage \\ 
\hline
P10 & Developer & 3-5 years & Male & 25-34 years old & Asian & Daily \\
P11 & Manager & 1-3 years & Male & 25-34 years old & Caucasian & Occasionally \\
P12 & Freelancer & 3-5 years & Male & 35-44 years old & Asian & Daily \\
P13 & Journalist & 1-3 years & Female & 18-24 years old & Hispanic & Weekly \\
P14 & Consultant & 3-5 years & Male & 25-34 years old & Caucasian & Daily \\
P15 & Product Designer & 3-5 years & Female & 25-34 years old & Caucasian & Weekly \\
P16 & Analyst & 5-10 years & Male & 35-44 years old & Caucasian, African American & Weekly \\
P17 & Scientist & 3-5 years & Female & 25-34 years old & Asian & Weekly \\
P18 & Scientist & 1-3 years & Female & 25-34 years old & African American & Weekly \\
P19 & Student & 5-10 years & Male & 25-34 years old & Caucasian & Only once or twice \\
P20 & Freelancer & > 10 years & Male & 45-54 years old & Caucasian & Occasionally \\
P21 & Student & 3-5 years & Male & 25-34 years old & Asian & Daily \\ \bottomrule
\end{tabular}%
}
\caption{Demographic and experience related information for participants in user study.}
\label{tab:participants}
\end{table*}

\section{Perception of practitioners toward \textsc{ChatGPT}'s utility}

While the VisGuides forum analysis findings are promising, they do not provide insights into how practitioners perceive the value of \textsc{ChatGPT}'s responses, prompting the need for further research. As a result, we conducted a comparative interview study involving feedback sessions with \textsc{ChatGPT} and \textsc{Human} experts. \added{Experts were chosen over anonymous users to ensure the quality of human responses and establish a higher benchmark for evaluating the limitations of \textsc{ChatGPT}. These experts also represent experienced colleagues who provide design feedback in professional settings~\cite{choi2023vislab,esteves2022learned}.} Our goal was to explore how data visualization practitioners evaluate \textsc{ChatGPT}'s effectiveness as a design companion, focusing on their experiences and perspectives on its design feedback.

\subsection{Recruitment}
We aimed to engage a diverse group of data visualization practitioners in our study. We recruited participants through academic mailing lists targeting students and scientists, as well as the Data Visualization Society's Slack channel~\cite{datavissociety}. Inclusion criteria required proficiency in English, experience in data visualization creation, and a willingness to share at least one of their data visualizations. Initially, 41 individuals responded to the recruitment survey, from which we selected 12 participants to take part in the study.

\subsection{Participants}
The study involved 12 participants (P10 to P21), creating a diverse group with varying professional backgrounds, experience levels, and familiarity with \textsc{ChatGPT} (see Table \ref{tab:participants}); we started participant codes from 10 to ensure that all IDs were consistently two digits long. Their roles encompassed a broad spectrum, including developers, managers, journalists, consultants, product designers, analysts, scientists, and students. Professional experience ranged from less than one year to over ten years in their respective fields. Furthermore, participants' engagement with \textsc{ChatGPT} varied, with some being daily users while others interacted with it only occasionally or as the need arose.

\subsection{Tasks \& Procedures}
Before each study session, participants were asked to prepare a visualization they were comfortable sharing, along with a list of relevant questions. These sessions were conducted using Zoom between the end of July and early August 2023. The entire study session took about 60 minutes. Participants were compensated with a \$50 Amazon gift card.

Each interview began with a brief introduction by the moderator and was divided into three segments: a visualization feedback session with \textsc{ChatGPT}, a similar session with \textsc{Human} expert(s), and an open-ended interview. \added{Our study employed a within-subjects design with counterbalancing to control for order effects.} Six participants initiated the process with the \textsc{ChatGPT} session, while the other six commenced with the \textsc{Human} expert session. After each feedback session, we administered a survey. In five sessions, two visualization experts provided feedback, while in the remaining seven sessions, we had one expert present. Both experts are current professors in the field of visualization.

During the \textsc{ChatGPT} feedback session, participants shared their screens with the moderator and presented their visualizations, accompanied by a brief explanation. The moderator introduced a predefined input format for \textsc{ChatGPT}, employing a role-playing structure~\cite{ihwan2023role}. Participants were guided to furnish details about the visualization's chart type, textual description, visual encodings, and any related questions. They input this information into \textsc{ChatGPT} while screen-sharing. Following \textsc{ChatGPT}'s response, participants could pose follow-up questions or request clarifications. Afterward, participants received a survey via the Zoom chat to gather feedback on their \textsc{ChatGPT} experience.

During the \textsc{Human} expert session, one or two visualization experts joined the Zoom meeting. Participants shared their screens and presented their visualization-related questions to these experts. The experts answered queries and provided additional feedback or further insights. Subsequently, following this session, the visualization experts left the meeting, and participants were presented with a Zoom survey to gather feedback about their experience.

In the final segment of the study, open-ended interviews were conducted. In these interviews, the moderator inquired about participants' experiences in both \textsc{ChatGPT} and \textsc{Human} feedback sessions, delving deeper into the rationale behind their survey responses. Furthermore, participants were prompted to share their perspectives on the possible integration of \textsc{ChatGPT} into their data visualization workflow, its constraints, and the future possibilities it might offer.

\subsection{Data \& Analysis Methods}
The final dataset from our user studies comprises participant questions (see Table \ref{tab:feedback-questions}) and responses, survey data, and transcripts from post interviews. Our primary focus for analysis was on the interview transcripts. Initially, two researchers independently examined three interview transcripts, identifying interesting quotes and assigning meaningful themes. Once we established a high degree of consistency between the researchers' findings, one of them proceeded to review the remaining transcripts. Ultimately, we compiled a categorized list of quotes based on initial higher-level themes (e.g., \textit{Where \textsc{Human} experts excel}). To further refine our analysis, we revisited these quotes and divided the themes into sub-level categories (e.g., \textit{Collaborative and natural conversations}).

\subsection{Post-Session Interview Analysis Results}

\begin{table*}[]
\resizebox{\textwidth}{!}{%
\begin{tabular}{lll}
\hline
P ID & Chart Type & Initial Questions \\ \hline
P10 & parallel coordinate map & Could you suggest an alternative way to visualize this? How can I make this visualization more engaging? \\
P11 & scatterplot & Does it make more sense to compare X or to compare Y? Would there be a better visualization type to show \_\_\_? \\
P12 & scatterplot & How can I improve this visualization in order to satisfy the customer? \\
P13 & scatterplot & What is the best way to show \_\_\_? what do you think about the current color scheme? \\
P14 & diverging stacked bar chart & How can we ensure the user looks at \_\_\_? How can I put these four bars next to each other in one visualization? \\
P15 & bubble cluster & How would I enable more properties to be seen? How would I let the user encode the strength of their preference? \\
P16 & interactive map & Which parts of the data visualization do you think are successful? Are there any areas you believe could be improved for better clarity and impact? \\
P17 & sankey chart & Is there a way to ascertain what visualization is most effective for communicating the data? How could \_\_\_ be more clearly displayed? \\
P18 & wav file visualization & Would it be visually overwhelming to use \_\_\_? Is this an appropriate visualization for demonstrating \_\_\_? \\
P19 & color heatmaps & Would another color map be more suitable? Would you have any suggestions as to how to indicate \_\_\_? \\
P20 & beeswarm, line chart & Does the special encoding for \_\_\_ make intuitive sense? Is it clear that \_\_\_ represents \_\_\_? \\
P21 & comparative line graph & How can I decrease the number of colors used in this visualization? Would this data be better represented if I \_\_\_? \\ \hline
\end{tabular}%
}
\caption{Initial questions from participants. These initial questions are altered to maintain the privacy of the participants. The original questions provided more contextual information and specificity.}
\label{tab:feedback-questions}
\end{table*}

\subsubsection{Participant Questions and Interaction Dynamics}


Four of the participants (P10, P15, P16, and P18) shared interactive visualizations. The other eight participants shared static visualizations. The type of visualization, as well as the questions they initially brought, can be seen in Table~\ref{tab:feedback-questions}. To safeguard participant privacy, we refrain from disclosing their visualizations or any elements of their questions that could potentially reveal information about participants. 
 
In the \textsc{ChatGPT} section of the interview, the majority of the participants asked \textsc{ChatGPT} more than one follow-up question. The follow-up questions to \textsc{ChatGPT} tended to be more specific and technical than the follow-up questions asked to the \textsc{Human} experts. For example, P11 asked \textsc{ChatGPT} \textit{``What are the pros and cons of having too many bubbles on the scatterplot from having a clear comparison?''}

On the other hand, the follow up questions to the \textsc{Human} experts tended to be less structured and more conversational.  
The participant and expert had a conversation and pulled out different aspects of the question along the way.
There was a lot of stopping to clarify what different elements of the visualization meant and how they played into the visualization. As a result, other factors not explicitly mentioned in the question often ended up getting brought up in these sessions with the \textsc{Human} experts.

\subsubsection{Post Survey Results}
\begin{figure*}
  \includegraphics[width=\textwidth]{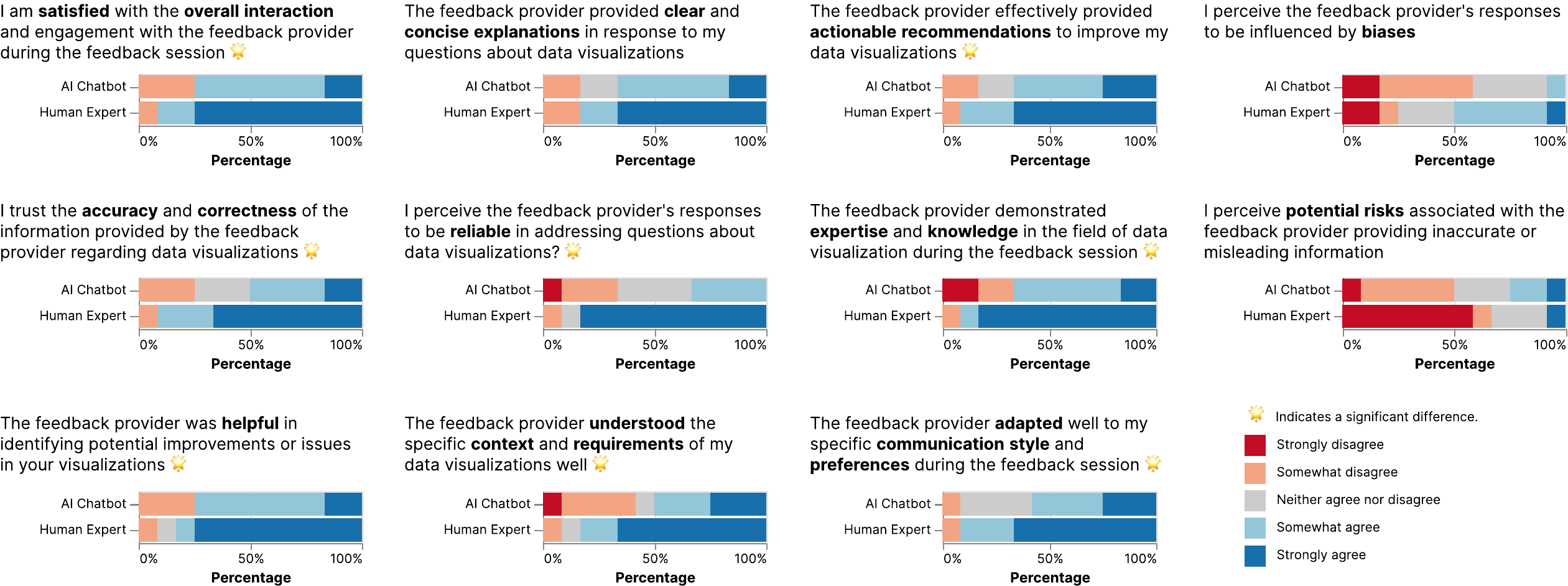}
  \caption{User preference for \textsc{Human} expert vs. \textsc{ChatGPT} responses in feedback sessions: The overarching pattern evident in these charts predominantly indicates a user preference for \textsc{Human} experts over \textsc{ChatGPT} in the context of feedback sessions.}
  
  \label{fig:survey-result}
\end{figure*}

In Figure \ref{fig:survey-result}, we present the outcomes of the experience surveys conducted following each feedback session. The overarching consensus among participants was a strong preference for \textsc{Human} experts. This preference was underscored by statistically significant distinctions, as determined through two-tailed Mann-Whitney U tests (1: strongly disagree to 5: strongly agree). Specifically, participants expressed significantly higher levels of satisfaction in interacting with \textsc{Human} experts ($U=113.5$, $p<0.05$), perceiving their responses as notably more accurate ($U=113.5$, $p<0.05$), helpful ($U=110.0$, $p<0.05$), reliable ($U=128.5$, $p<0.05$), and adaptable ($U=105.5$, $p<0.05$) to their preferences.

Furthermore, respondents indicated that \textsc{Human} experts exhibited a significantly deeper understanding of the context and requirements of their queries. \textsc{Human} experts were also acknowledged for their expertise in data visualization, along with their ability to offer actionable recommendations. It is worth noting that no significant differences were observed in responses to questions related to the clarity and conciseness of explanations, concerns about potential biases, or perceived risks of receiving misleading information.

\subsubsection{Where \textsc{Human} Experts Excel}

Participants were generally more satisfied with \textsc{Human} experts. They mentioned a variety of reasons for such opinions. The participants highlighted several key themes in their responses, shedding light on the strengths of \textsc{Human} experts in the context of data visualization guidance.

\paragraph{Bespoke and focused feedback of \textsc{Human} experts}
First of all, they liked \textsc{Human} experts' \textbf{ability to offer tailored recommendations} that were closely aligned with the specific visuals in question (P10, P11, P15, P17 - P21). P17 exemplified this sentiment by stating, \textit{``[...] recommendations are a lot more tailored and specific in that way, versus just like being tossed a list of potential tools we can look.''}. P11 further elaborated on this point by highlighting that \textsc{Human} experts can envision various options and present the ones they deem most valuable. They also noted that \textsc{Human} experts could effectively \textbf{understand the context of visuals} by directly observing them (P11, P17 - P19). P19 expressed this sentiment by stating, \textit{``I felt the feedback was obviously more grounded to the visualization at hand. Because again, they could see it.''} Similarly, P18 commented, \textit{``As he's looking at the pictures, he knows the information I had access to and how that translates.''} Furthermore, participants acknowledged that \textsc{Human} experts excelled in \textbf{staying focused on the specific problem at hand}. For instance, P15 articulated, \textit{``[the expert] was able to talk about improvements within scope [...] without redesigning it entirely.''} Similarly, P10 commented, \textit{``The ones that [the expert] gave were within the scope of refining existing visualizations.''}

\paragraph{Collaborative and natural conversations}
Participants consistently expressed the value of more fluid conversations when interacting with \textsc{Human} experts (P10, P11, P15, P17, P19, P20). They highlighted that \textsc{Human} responses \textbf{felt more free-flowing}, contributing to a sense of genuine interaction (P10, P17). Participants also noted that \textsc{Human}-expert interactions provided a \textbf{collaborative and interactive} experience (P11, P15, P19, P20). For instance, P11 highlighted an efficient turn-taking dynamic by saying, \textit{``the ability to ask follow-up questions and clarify I think was easier to do within the \textsc{Human} interaction context.''} Others similarly emphasized the convenience of seeking clarifications without the need for precise phrasing or additional context adjustments (P11, P15).

\paragraph{Enriched insights through lived experience}
Participants felt that \textsc{Human} experts' responses were often \textbf{rooted in their experience} and education (P16).
They valued the ability of \textsc{Human} experts to explain the underlying rationale, promoting a more actionable approach to implementing the suggestions (P10). Furthermore, participants appreciated the \textsc{Human} experts' ability to think beyond the obvious; as P13 noted, \textit{``I feel like I was definitely more satisfied with the \textsc{Human} experts, [...] they just started thinking so many things beyond anything I could have think of.''} Moreover, participants conveyed that interactions with \textsc{Human} experts \textbf{felt personal and supportive}. One participant reflected, \textit{``they were really walking me through the different things [...] help you grow as a professional [...] really pushing me.''}, while another participant recognized the \textbf{proactiveness} of \textsc{Human} experts by saying \textit{``even if I didn't ask question, they kind of brought it up.''}

\subsubsection{Where \textsc{ChatGPT} excels}
Participants identified the strengths of \textsc{ChatGPT} in contrast to \textsc{Human} experts.

\paragraph{Brainstorming and ideation}
\textsc{ChatGPT}'s role as a \textbf{creative catalyst} emerged prominently during participant discussions, underlining its capacity to spark innovative ideas  (P11, P15, P18).  Participants like P18 found unexpected and impressive suggestions, stating, \textit{``It did have some cool ideas, I didn't expect this from \textsc{ChatGPT}. [...] Pretty good.''}, while P11 commented, \textit{``I think it's really interesting to try and kind of break outside the box and get more creative... GPT has the potential to disrupt that workflow in a positive way.''} Likewise, P15 also acknowledged that the tool may not always provide precise solutions but rather serve as a springboard for creative exploration. P15 contrasts this to \textsc{Human} experts by saying, \textit{``[expert] feedback was less brainstorming and more actual critique.''}

\paragraph{Broad knowledge base} 
In a similar vein to its brainstorming capability, participants also highlighted \textsc{ChatGPT}'s capacity to serve as a \textbf{vast repository of knowledge}, offering a wide array of ideas (P10, P11, P15). P10 stated, \textit{``it really does kind of feel like a really fancy kind of search engine, where it is kinda, I give it a problem. And it's like, oh, like, this might be a solution.''}, while P15 said, \textit{``great place to go to start [...] researching or come up with ideas you hadn't thought of.''} P11 similarly emphasized that \textsc{ChatGPT} excels at showcasing a range of data visualization options, making it a valuable resource during the initial stages of projects. P11 elaborated on the tool's potential, envisioning a scenario where metadata about a dataset could be fed to \textsc{ChatGPT} to receive chart-type recommendations. However, P10 emphasized that a \textbf{critical eye is needed} to discern the relevant and valuable insights from its outputs; \textit{``I asked to give 20 ideas and I'm [...] experienced enough [...] to [...] know that, like, 16 of them are just nonsense. The other four things I might not have thought of.''}

\paragraph{Time saved via rapid understanding and response}
Participants appreciated \textsc{ChatGPT}'s ability to \textbf{quickly grasp and provide information} (P10, P11, P13). P10 said \textsc{ChatGPT} was able to quickly grasp unfamiliar concepts auxiliary to visualization (e.g., explaining Wordle to \textsc{ChatGPT}), while P11 and P13 felt \textsc{ChatGPT} provided faster and more comprehensive lists of options, especially for alternative chart types, when compared to the responses from \textsc{Human} experts. On a related note, P11 said that \textsc{Human} interactions can sometimes involve a lengthier back-and-forth process to arrive at the right question.

\paragraph{Soft attitude and neutral perspective}
Other participants noted \textsc{ChatGPT}'s behavioral traits. For example, P19 pointed out that \textsc{ChatGPT} often assumed a \textbf{gentle and agreeable} attitude by acknowledging concerns without offering specific directions, which led to a perception of excessive alignment. P19 also noted that \textsc{ChatGPT} demonstrated a \textbf{neutral stance}, presenting a variety of viewpoints without favoring any particular side. This contrasted with \textsc{Human} experts who might convey their own biases, thereby making \textsc{ChatGPT}'s approach seem more balanced.

\subsubsection{Where \textsc{ChatGPT} falls short}
Participants also discussed \textsc{ChatGPT}'s several shortcomings in comparison to \textsc{Human} experts.

\paragraph{Lack of depth in responses}
Participant feedback indicated a notable lack of depth in \textsc{ChatGPT}'s responses (P13, P15, P19). For instance, P19 expressed that \textsc{ChatGPT}'s advice seemed to lack actionable insights and depth that \textsc{Human} experts possess. This sentiment was echoed by P15, who noted that while \textsc{ChatGPT}'s answers were broad in knowledge, they lacked depth. P15 further elaborated that \textsc{ChatGPT} often suggested radical changes (i.e., diverse ideas), while \textsc{Human} experts provided improvements that were more aligned with refining existing visualizations. P19 similarly commented that \textsc{ChatGPT}'s ideas were not considering the specific visualization. Participants acknowledged this contrast as a trade-off (P15, P17), emphasizing that the desired level of response detail and granularity from either \textsc{ChatGPT} or \textsc{Human} experts depends on user needs.

\paragraph{Generic advice, misalignment, and lack of critical thinking}

Participants expressed that \textsc{ChatGPT}'s feedback often \textbf{felt generic and lacked contextual depth} (P12, P16, P20). P20 expressed that \textsc{ChatGPT}'s \textit{``knowledge felt [...] not necessarily backed up by experience.''} P16 and P12 similarly noted that \textsc{ChatGPT}'s responses were not necessarily wrong but were overly generic, failing to consider the actual outcomes of the provided recommendations. P16 further commented that they \textit{``felt the areas of improvement were good suggestions. But it didn't feel like you have to do this.''}

Several participants highlighted instances of \textbf{misalignment and a lack of specificity} in \textsc{ChatGPT}'s recommendations (P11, P16, P19). P19 pointed out an example where \textsc{ChatGPT} suggested color maps that might create confusion rather than improve visualizations. This lack of alignment was further emphasized by P11, who noted a communication hurdle hindering the establishment of a common understanding and precise feedback.

Participants highlighted \textbf{concerns related to critical thinking} (P13, P16). P16 noted, \textit{``whereas the \textsc{ChatGPT} felt more like it was telling me what I wanted to hear if that makes sense.''} P16 further elaborated that \textsc{ChatGPT}'s responses sometimes resembled textbook information rather than offering insightful suggestions on visualization. P13 shared a similar sentiment, stating, \textit{``And I feel like the issue with \textsc{ChatGPT} typically is that it only responds to your question. So if you cannot think of the question, it will not give you the answer.''}

\paragraph{Efforts required for fluid conversation}
Participants highlighted the \textbf{challenges in having a fluid and interactive conversation} with \textsc{ChatGPT} (P10, P13, P16, P17, P20). For instance, P10 explained, \textit{``... kind of hoping that one of them would be what I wanted.''}. P17 expressed a similar sentiment, stating, \textit{``even though it's still conversational, but it still kind of feels a little bit static sometimes where you're just typing text into a text box.''} P16 further emphasized the difference in interaction dynamics, stating, \textit{``I really liked listening to the \textsc{Human} experts go back and forth, and talk about different aspects... whereas \textsc{ChatGPT} just fed that to me, and that was it.''} P13 noted that \textsc{ChatGPT}'s responses often required iterations and follow-ups, P10 found that \textsc{ChatGPT}'s insights became more specific and interesting when more constraints were provided.

\subsubsection{Opinions on Trust and Reliability with \textsc{ChatGPT}}

Participants had \textbf{varied opinions on the trustworthiness and reliability} of \textsc{ChatGPT}'s responses, with some expressing low trust (P10, 18, P20). P20 mentioned the limitations in \textsc{ChatGPT}'s understanding of visuals, which relied heavily on the accuracy of participants' descriptions. P18 pointed out the potential for fabricated information due to the algorithmic nature of the responses. P10 echoed this sentiment, indicating that \textsc{ChatGPT} often sounded too confident and authoritative, which could mislead users into placing unwarranted trust. On the other hand, P17 said the two are relatively the same, highlighting that the key distinction lies in how information is delivered.

 Several other participants pointed to the need for \textbf{due diligence in evaluating \textsc{ChatGPT}'s suggestions}, especially when the user lacks domain expertise (P10, P15). For instance, P10 commented: \textit{``If I was in a field that I wasn't familiar with, I think it'd be really easy to get fooled by it.''} P15 drew a parallel with the early inception of Wikipedia, stating, \textit{`` it was like early days of Wikipedia, I was taught like, never cite Wikipedia, it could be wrong. And I think the same thing is true of something like a large language model.''}

 They also made comparisons to \textsc{Human} experts in terms of trust and authority (P13, P16, P18, P21). They explained how the \textbf{credentials of \textsc{Human} respondents} influenced their perception (P18, P21), e.g., \textit{``Well, the \textsc{Human} respondent is a professor in visualization.''}---P18. Others perceived \textsc{Human} experts as sounding knowledgeable and educational (P13, P16). In contrast to these, participants also shared their perceptions of \textbf{expertise and self-awareness in AI}. P16 said, \textit{``Didn't feel like that the \textsc{ChatGPT} response was incorrect ... trusted it, but not nearly as much as the \textsc{Human}s.''}, while P20 expressed, \textit{``\textsc{ChatGPT} doesn't know when it's wrong.''}

\subsubsection{Limitations and Opportunities with \textsc{ChatGPT}}

Participants expressed an \textbf{optimistic outlook} regarding the future potential of \textsc{ChatGPT}, despite acknowledging its current challenges (P10, P12, P15, P16). They discussed aspects they would like to see improved in \textsc{ChatGPT} in the future.

\paragraph{Ability to convey complex visualizations}
Participants emphasized the significance of enabling \textsc{ChatGPT} to comprehend and interpret complex visualizations, highlighting the \textbf{limitations of text-based communication} when dealing with intricate design problems (P10, P11, P13, P15 - 21). P15 acknowledged the inherent loss of information when translating between visual and written mediums, while P10 and P11 conveyed the complexity of describing interactive dashboards and multimedia visualizations via text. P10 also mentioned video recordings that could convey the interactivity of designs.

\paragraph{Ability to generate visualizations based on feedback}

Several participants pondered the prospect of \textsc{ChatGPT} being able to not only analyze input information but also \textbf{generate relevant visualizations} (P18, P19, P21). For example, P18 proposed the generation or modification of images based on textual descriptions. P19 noted the convenience of sharing underlying code to accompany visual content, emphasizing its value in effectively communicating complex ideas instead of relying solely on text descriptions. Similarly, P21 discussed the concept of inputting data and posing questions to explore the space of chart design. P13 envisioned the integration of \textsc{ChatGPT} within existing tools, enhancing workflow efficiency and facilitating a more cohesive user experience.

\paragraph{Facilitating Fluid and Truthful Conversations}

Participants expressed the need for \textsc{ChatGPT} to initiate and support more fluid conversations to enhance its usability and value (P11, P16, P21). P16 highlighted the need for AI to \textbf{prompt users for further details} about the visualization; \textit{``I would really like it if it were to ask me questions to kind of gauge my thought process.''} Similarly, P11 noted the \textbf{value of follow-up questions} in driving valuable insights, while \textit{``it was harder to envision what a follow-up question to that system would be.''} On the other hand, P21 raised concerns about the possibility of \textsc{ChatGPT} providing false or hallucinated information.

\paragraph{The indispensable role of \textsc{Human} feedback}
The consensus among participants highlighted the \textbf{enduring value of \textsc{Human} feedback} in evaluating and improving visualizations (P11, P16, P21). P16 commented, \textit{``I think there's always, at least in my lifetime, going to be a need for \textsc{Human} interpretation. And the \textsc{Human} experts' sessions, for me, reinforce that they offer a lot of really good insight.''} P11 further expressed that despite \textsc{ChatGPT}'s capabilities, the feedback from actual users remains incomparable, as it encompasses a deeper level of insight into user experience.  Additionally, P11 discussed ideas of experimenting with different \textbf{personas for hypothetical user interactions}.

\paragraph{Issues with privacy and sensitive data}

Participants in the interviews expressed significant concerns regarding the privacy and handling of sensitive data by \textsc{ChatGPT} (P15, P16, P20, P21). They emphasized that sharing visuals containing private or \textbf{sensitive information would be uncomfortable} and unlikely unless the data was from a public source or there was a guarantee that it would not be shared publicly. P15 shared an illustrative example, stating, \textit{``I met a company that I had to sign an NDA with, and we are not allowed to use \textsc{ChatGPT} with anything that might be considered confidential information because that information goes onto OpenAI servers and will live there for eternity.''}



\subsection{Takeaways}

The findings from our study underscore the distinct strengths and limitations of \textsc{ChatGPT} in comparison to \textsc{Human} experts. The clear preference for \textsc{Human} experts among participants highlights their superior ability to provide tailored, context-specific advice and engage in more interactive and collaborative conversations. This preference is rooted in the \textsc{Human} experts' deep understanding of visualization context and their capacity for nuanced, bespoke feedback. \added{The feedback provided by \textsc{Human} experts in these sessions closely paralleled the high-scoring Human responses observed in the Visguides.}

On the other hand, \textsc{ChatGPT} shows promise as a creative catalyst and a broad knowledge base, offering rapid responses and a diverse array of ideas. However, its limitations are evident in the lack of depth and specificity in responses, challenges in fluid conversation, and concerns about perceived trustworthiness~\cite{yin2019understanding,toreini2020relationship,drozdal2020trust,yang2020visual,chatzimparmpas2020state} and handling sensitive data~\cite{liu2021machine}. Several participants expressed concerns about needing expertise to assess the validity and relevance of potentially inaccurate suggestions from \textsc{ChatGPT}. Nonetheless, they remained optimistic about its future potential, anticipating continued improvements and increasing value.

\section{Discussions}


\removed{Both studies consistently highlighted similar strengths and weaknesses of \textsc{ChatGPT} compared to \textsc{Human} counterparts. 
A notable difference was that participants generally showed a stronger preference for \textsc{Human} experts in the second study. This disparity might be attributed to several factors. One reason could be that while any member of the VisGuides community can respond to questions on the forum, the feedback sessions were consistently answered by experienced visualization experts. In generall, the responses from \textsc{Human} experts in these sessions closely paralleled the high-scoring \textsc{Human} responses observed in the Visguides. Moreover, human-to-human interaction might have also played a role. For instance, \textsc{Human} experts ensured that they did not overlook any aspects of participants' questions, benefiting from real-time conversations. Overall, the second study might demonstrate that human perception of usefulness might extend beyond just response quality.
}

\subsection{Limitations of our studies and future opportunities}

While our studies were designed and conducted with considerable care, there exist certain limitations. First, the absence of demographic and expertise data for the \textsc{Human} respondents in the VisGuides forum makes direct comparisons with the \textsc{Human} experts challenging. \added{However, the human baselines in both studies represent distinct real-world scenarios: receiving feedback from anonymous online platforms or from experienced colleagues in the workplace.} \removed{Also, we relied on the experiential opinions of participants to assess the feedback sessions, providing more direct and accurate evaluations.} Also, due to the dynamic nature of the feedback sessions in the second study, we were not able to apply the static evaluation metrics used in the VisGuides. Despite this limitation, our findings were highly consistent across the two studies.

We used a fixed version of \textsc{ChatGPT}\removed{, a fine-tuned iteration of GPT 3.5}. \removed{As with many other new technologies, LLMs are continuously advancing.} \textsc{ChatGPT} and other LLM-based chatbots are evolving to incorporate more recent data and support new features, including the ability to handle more nuanced language or to understand and generate images, although some of these features are only available to paid users. As a result, responses to the questions in our study can vary among newer LLMs. \added{Likewise, the recently released advanced voice mode could help narrow the interaction modality gap identified in the second study, potentially improving the fluidity of conversations. That said, it still falls short of achieving human-like naturalness, such as exhibiting proactive questioning and avoiding passive behavior.} \removed{: they can be different, better, or sometimes even worse in certain cases.} \added{
However, our study results provide a valuable foundation for future research, offering both datasets, metrics, and methodologies for evaluating LLM responses in the context of visualization design feedback.}



\subsection{Towards automatic evaluation of LLM feedback capabilities}
\label{sec:llm-based-eval}
\added{Facilitating ongoing monitoring of LLMs' design feedback capabilities would be beneficial to ensure that both the development of LLMs and the systems built upon them remain well-grounded. The current reliance on human intervention to evaluate the quality of responses is inefficient. Recently, LLMs have demonstrated remarkable capabilities in evaluating natural language generation, showing greater alignment with human evaluations compared to traditional similarity-based or embedding-based metrics such as BLEU or BERTScore~\cite{zheng2023judging,liu2023gpteval,wang2023chatgpt,fu2023gptscore,kim2023prometheus}.}

\begin{table}[]
    \begin{tabular}{@{}llccccl@{}}
    \toprule
    \textbf{Metric}     & \textbf{Response type} & \textbf{Mean diff} & \textbf{$\tau$} & \textbf{$\rho$} & \textbf{Gwet's AC1} & \textbf{Score dist} \\ \midrule
    
    \multirow{2}{*}{Depth} & Human  & -0.563     & 0.558   & 0.631   & 0.208       
    & \begin{tikzpicture}
    \begin{axis}[axis lines=left, width=3.5cm, height=1.85cm, ybar, enlarge x limits=0.15, axis line style= -, ytick=\empty, xtick=\empty]
    \addplot coordinates {(1,3.00) (2,3.82) (3,6.00) (4,4.50) (5,4.77)};
    \end{axis}
    \end{tikzpicture}    \\
   & ChatGPT& 0.542      & 0.215   & 0.228   & 0.29
   & \begin{tikzpicture}
    \begin{axis}[axis lines=left, width=3.5cm, height=1.85cm, ybar, enlarge x limits=0.15, axis line style= -, ytick=\empty, xtick=\empty]
    \addplot coordinates {(1,3.00) (2,3.13) (3,6.00) (4,4.79) (5,3.40)};
    \end{axis}
    \end{tikzpicture}      \\ \midrule
   
    \multirow{2}{*}{Breadth}       & Human  & -0.24      & 0.553   & 0.62    & 0.424       
    & \begin{tikzpicture}
    \begin{axis}[axis lines=left, width=3.5cm, height=1.85cm, ybar, enlarge x limits=0.15, axis line style= -, ytick=\empty, xtick=\empty]
    \addplot coordinates {(1,3.00) (2,4.75) (3,6.00) (4,4.75) (5,4.75)};
    \end{axis}
    \end{tikzpicture}    \\
   & ChatGPT& -0.083     & 0.522   & 0.537   & 0.713       & \begin{tikzpicture}
    \begin{axis}[axis lines=left, width=3.5cm, height=1.85cm, ybar, enlarge x limits=0.15, axis line style= -, ytick=\empty, xtick=\empty]
    \addplot coordinates {(1,3.00) (2,3.00) (3,3.29) (4,4.06) (5,6.00)};
    \end{axis}
    \end{tikzpicture}      \\ \midrule
    
    \multirow{2}{*}{Clarity}       & Human  & -0.792     & 0.372   & 0.395   & 0.057       
    & \begin{tikzpicture}
    \begin{axis}[axis lines=left, width=3.5cm, height=1.85cm, ybar, enlarge x limits=0.15, axis line style= -, ytick=\empty, xtick=\empty]
    \addplot coordinates {(1,3.00) (2,3.11) (3,4.11) (4,4.79) (5,6.00)};
    \end{axis}
    \end{tikzpicture}    \\
   & ChatGPT& -0.01      & 0.517   & 0.52    & 0.901       & \begin{tikzpicture}
    \begin{axis}[axis lines=left, width=3.5cm, height=1.85cm, ybar, enlarge x limits=0.15, axis line style= -, ytick=\empty, xtick=\empty]
    \addplot coordinates {(1,3.00) (2,3.03) (3,3.03) (4,3.22) (5,6.00)};
    \end{axis}
    \end{tikzpicture}      \\ \midrule
    \multirow{2}{*}{Topicality}    & Human  & -0.771     & 0.366   & 0.396   & 0.208       
    & \begin{tikzpicture}
    \begin{axis}[axis lines=left, width=3.5cm, height=1.85cm, ybar, enlarge x limits=0.15, axis line style= -, ytick=\empty, xtick=\empty]
    \addplot coordinates {(1,3.00) (2,3.00) (3,3.24) (4,3.65) (5,6.00)};
    \end{axis}
    \end{tikzpicture}   \\
   & ChatGPT& 0.063      & -0.029  & -0.029  & 0.915       & \begin{tikzpicture}
    \begin{axis}[axis lines=left, width=3.5cm, height=1.85cm, ybar, enlarge x limits=0.15, axis line style= -, ytick=\empty, xtick=\empty]
    \addplot coordinates {(1,3.00) (2,3.00) (3,3.08) (4,3.25) (5,6.00)};
    \end{axis}
    \end{tikzpicture}      \\ \midrule
   
    \multirow{2}{*}{Actionability} & Human  & -0.928     & 0.47    & 0.525   & 0.167       
    & \begin{tikzpicture}
    \begin{axis}[axis lines=left, width=3.5cm, height=1.85cm, ybar, enlarge x limits=0.15, axis line style= -, ytick=\empty, xtick=\empty]
    \addplot coordinates {(1,3.00) (2,3.20) (3,5.20) (4,6.00) (5,5.30)};
    \end{axis}
    \end{tikzpicture}    \\
   & ChatGPT& 0.406      & 0.466   & 0.492   & 0.422       & \begin{tikzpicture}
    \begin{axis}[axis lines=left, width=3.5cm, height=1.85cm, ybar, enlarge x limits=0.15, axis line style= -, ytick=\empty, xtick=\empty]
    \addplot coordinates {(1,3.00) (2,3.05) (3,3.75) (4,6.00) (5,3.75)};
    \end{axis}
    \end{tikzpicture}      \\ \midrule
    
    \multirow{2}{*}{Coverage}      & Human  & -0.87      & 0.41    & 0.457   & 0.097       
    & \begin{tikzpicture}
    \begin{axis}[axis lines=left, width=3.5cm, height=1.85cm, ybar, enlarge x limits=0.15, axis line style= -, ytick=\empty, xtick=\empty]
    \addplot coordinates {(1,3.00) (2,3.29) (3,3.29) (4,3.88) (5,6.00)};
    \end{axis}
    \end{tikzpicture}    \\
   & ChatGPT& 0.01       & 0.424   & 0.424   & 0.925       & \begin{tikzpicture}
    \begin{axis}[axis lines=left, width=3.5cm, height=1.85cm, ybar, enlarge x limits=0.15, axis line style= -, ytick=\empty, xtick=\empty]
    \addplot coordinates {(1,3.00) (2,3.00) (3,3.06) (4,3.22) (5,6.00)};
    \end{axis}
    \end{tikzpicture}      \\ \midrule
    
    \multirow{3}{*}{All}   & All    & -0.27      & 0.404   & 0.433   & 0.444       
    & \begin{tikzpicture}
    \begin{axis}[axis lines=left, width=3.5cm, height=1.85cm, ybar, enlarge x limits=0.15, axis line style= -, ytick=\empty, xtick=\empty]
    \addplot coordinates {(1,3.029)(2,3.137)(3,3.540)(4,3.684)(5,4.610)};
    \end{axis}
    \end{tikzpicture}      \\
   & Human  & -0.694     & 0.455   & 0.504   & 0.194       & \begin{tikzpicture}
    \begin{axis}[axis lines=left, width=3.5cm, height=1.85cm, ybar, enlarge x limits=0.15, axis line style= -, ytick=\empty, xtick=\empty]
    \addplot coordinates {(1,3.057)(2,3.251)(3,3.662)(4,3.706)(5,4.324)};
    \end{axis}
    \end{tikzpicture}      \\
   & ChatGPT& 0.154      & 0.353   & 0.362   & 0.694       & \begin{tikzpicture}
    \begin{axis}[axis lines=left, width=3.5cm, height=1.85cm, ybar, enlarge x limits=0.15, axis line style= -, ytick=\empty, xtick=\empty]
    \addplot coordinates {(1,3.000)(2,3.022)(3,3.419)(4,3.662)(5,4.897)};
    \end{axis}
    \end{tikzpicture}      \\ \bottomrule
    \end{tabular}
    \caption{\added{The experiment results of the LLM-based automatic evaluation. The alignment between scores from LLM and human evaluators was derived from mean difference (Human score - LLM score), and the Kendall's tau ($\tau$), Spearman ($\rho$), and Gwet's AC1 (AC1) correlation coefficients. The histograms show the score distribution ranging from 1 (left) to 5 (right) for each metric.}}
    \label{tab:llm-evaluation}
    \end{table}

\added{We tested the viability of this idea by developing an evaluation prompt (see below) and used it to assess both \textsc{Human} and \textsc{ChatGPT 3.5}'s responses based on the six evaluation metrics outlined in Table~\ref{tab:rubrics}.}

\added{In the experiment, we employed the in-context learning method \cite{brown2020language, li2024generation} using \textbf{GPT-4o-mini via API calls in December 2024}, leveraging the LLM's few-shot learning capabilities to generalize evaluation tasks from the provided examples. Based on these settings, the evaluation task was conducted in the following steps: First, the Visguides dataset was split into a few-shot development set (20\%) and a test set (80\%). Second, for each test case and a metric, $n$ examples were selected from the few-shot development set, where we settled on $n=5$ after testing values within a range between 1 to 8 given the limit of context lengths in the GPT-4o-mini. The process for the example selection was based on two principles: 1) Score pooling for extracting diverse examples: To ensure a diverse pool of examples ranging from low to high quality, the examples in the few-shot development set were grouped into bins for unique scores from 1 to 5 based on their scores for the target metric. 2) Embedding similarity for retrieving relevant examples: Within each bin, the most similar examples based on embedding similarity based on Sentence Transformers \cite{devlin2018bert} were selected to help guide the model to choose the examples that are most semantically close to the input query, effectively ``picking the best match'' from the limited example set. The number of examples was determined proportionally to the size of the corresponding bin. The selected examples were incorporated into the prompt and provided to GPT-4o-mini, released in August 2024, as presented below:}

\begin{small}
\begin{verbatim}
You are an expert evaluator specializing in data visualization assessments. 
Your task is to evaluate responses to visualization-related questions as closely 
aligned with given examples, focusing specifically on the {metric} metric. 
Here are specific guidelines for {metric}:
    Before evaluating, carefully consider:
    1. The specific visualization requirements in the question
    2. How well the response addresses those requirements
    3. The technical accuracy and appropriateness of the visualization approach
    4. The practicality and implementability of the solution
   {metric-specific evaluation rubric}
   {metric-specific evaluation guideline}
   
Here are some example evaluations:
   {few-shot examples}

Now evaluate this {response type} response based on the following metric:
    Question: {question}
    Response to evaluate: {response}

Evaluate the response on {metric} using a scale of 1-5 (1 being lowest, 5 being 
highest). Your job is to make your evaluation as closely aligned as the given examples,
calibrating the score closely to the examples.
    
Metric definition:
- {metric}: {metric definition}

For the criterion, provide:
1. A numerical score (1-5)
2. A brief justification for the score
3. Specific examples from the response

Your response MUST be a valid JSON object with this exact structure:
{
    "evaluation": {
        "{metric}": {
            "score": X, 
            "justification": "...", 
            "examples": "..."
        }
    },
}

\end{verbatim}
\end{small}

\added{In the prompt, the metric-specific information such as evaluation rubric and guideline (Table \ref{tab:eval-rubrics} and \ref{tab:eval-guidelines}), few-shot examples, metric definition (Table \ref{tab:rubrics}), and response type (\textsc{Human} or \textsc{ChatGPT}) are integrated for each test case accordingly.}

\added{We then compare the output scores to those provided by human evaluators and calculate the alignment using agreement measures and correlation coefficients. In our experiment, we use Kendall’s tau ($\tau$), Spearman ($\rho$), and Gwet’s AC1 coefficient as underlying measures. While Kendall’s tau was selected as one of the most widely used metrics, several studies \cite{hou2024mitigating, liu2023gpteval} have pointed out that this measure can lead to biased results when the distribution of the underlying scores is severely skewed, failing to capture the true alignment. This issue is particularly relevant in our case for metrics such as clarity, coverage, and topicality, especially for \textsc{ChatGPT}'s responses, where scores are skewed toward higher values (e.g., 4 or 5). To address this, Gwet’s AC1 \cite{gwet2001handbook}, which has been proposed and increasingly adopted to mitigate biased judgments, was used to complement Kendall's tau.}

\added{Table \ref{tab:llm-evaluation} presents the outcomes of these measures for each evaluation metric. Overall, the results show that the LLM evaluator’s scores are fairly aligned with those of humans ($\tau=0.404$, $\rho=0.433$) obtained across all metrics, given a well-known interpretation of Kendall's tau and Spearman coefficient \cite{akoglu2018user, dancey2007statistics} ($\pm$0.4-0.6: moderate agreement). Specifically, the alignment was better captured for both \textsc{Human} and \textsc{ChatGPT} responses in the breadth ($\tau=0.553, \rho=0.62$; $\tau=0.522, \rho=0.637$) and actionability metric ($\tau=0.47, \rho=0.525$; $\tau=0.466, \rho=0.492$), and in the depth metric for \textsc{Human} responses ($\tau=0.558, \rho=0.631$). For metrics with highly skewed score distributions, such as topicality, Kendall’s tau coefficients were relatively low for ChatGPT responses. In these cases, Gwet’s AC1 coefficient provided a more nuanced perspective, demonstrating a high alignment ($\tau=-0.029, \rho=-0.029, AC1=0.915$). While our experiment shows promising capabilities of automatic evaluation using LLMs, it is worth to further investigate refining heuristics and developing prompts, for example, making prompts targeted to response types or providing clearer metric descriptions, in order to improve the alignment.}

\section{Design considerations for LLM-based feedback systems}
Our study results indicate the feasibility of developing an advanced design feedback system using LLMs. Drawing from the insights gained, we address several key design considerations for constructing such a system by addressing the strengths and limitations we identified in the study.

\paragraph{Leverage the knowledge base for creative exploration}
It was evident that the broad knowledge base of LLMs is advantageous for rapidly exploring diverse ideas or sparking new ideas for participants. However, one issue we observed in the study was that this discovery process was mostly ad-hoc and spontaneous, resulting in a hit-or-miss experience. Future feedback systems have the potential to make this process more systematic and supportive of structured ideation. This could be achieved by providing prompts or guiding questions that steer exploration toward relevant and productive paths. For example, a recent study provided insights into a structured exploration of the design space by generating dimensions to guide response generation~\cite{suh2023structured}.

\paragraph{Support nuanced understanding of design context}
One of the most frequent complaints was the lack of a nuanced understanding of the problem context in the question. LLMs with enhanced image understanding capability will alleviate this issue but will not be enough.  Future systems might consider allowing visual prompting in which users can annotate areas of interest. Participants also mentioned generating image responses that can help users envision how to address their design concerns; i.e., these responses could also be annotated to highlight improvements. Moreover, communicating design issues about interactive data visualizations poses interesting research challenges for future systems.

\paragraph{Facilitate fluid and proactive engagements}
Our participants favored seamless and dynamic discussions with \textsc{Human} experts. \textsc{ChatGPT} is currently trained to align with users' intentions~\cite{ouyang2022training}, often resulting in passive and affirmative behavior. Future feedback systems can aim to be more proactive, especially for novice users who may struggle to articulate their design concerns, such as by asking clarification questions or suggesting related design questions. Participant comments indicate that this approach may involve a trade-off, such as potentially offering irrelevant feedback when not needed, which can result in a suboptimal user experience. Therefore, it would be beneficial for users to have control over the level of proactiveness.

\paragraph{Improve trustworthiness of design suggestions}

Interestingly, although there is no guarantee that \textsc{Human} experts' responses are more truthful, participants were more concerned about \textsc{ChatGPT}'s credibility. Future systems take inspiration from the behavior of \textsc{Human} experts in our study. Having a fluid back-and-forth conversation to build a common ground was one thing, but they also often accompanied their responses with rationalized explanations and supplemented supporting empirical evidence. In a similar vein, participants were also uneasy about not being able to verify \textsc{ChatGPT} responses especially when they lacked relevant knowledge. Providing application examples, as they do in practice~\cite{choi2023vislab,kang2018paragon} might alleviate this concern. These abilities might require more than just prompt engineering, such as augmented retrieval~\cite{lewis2020retrieval}.

\paragraph{Integrating into a practitioner's workflow}
Several participants mentioned that they already use \textsc{ChatGPT} for tasks related to writing or debugging code (P10, P12, P13, P16, P18). But, data visualization practice rests on two pillars: design and implementation. Our study sheds light on \textsc{ChatGPT}'s capability in terms of design knowledge. As P13 suggests, the integration of a \textsc{ChatGPT}-like assistant within current tools would be highly beneficial. Such a design assistant could suggest appropriate chart types (P11, P21) or provide rationales or critiques for generated visualizations.




\paragraph{\added{Toward a better design knowledge education}}
Some participants noted \textsc{ChatGPT}'s potential for education (P20, P21), such as assisting with understanding unfamiliar charts or aiding in writing educational blogs on data visualization. A recent paper identifies challenges in visualization education~\cite{bach2023challenges}, highlighting AI as a valuable tool for formal and informal learning contexts. However, concerns include the consistency and quality of AI-generated content, as differing responses to students could be problematic. There's also the risk of students over-relying on AI instead of engaging critically with visualizations, e.g., through peer feedback. Future AI systems in education may act as personalized tutors, tracking progress, guiding learning goals, and fostering both comprehension and generation of visualizations.

\section{Conclusion}

In this study, we explored the potential of ChatGPT as a design companion, focusing on its capacity to provide valuable visualization insights to practitioners. Our findings reveal that ChatGPT
capitalizes on a vast knowledge repository to generate diverse and imaginative suggestions. While limitations like the contextual depth of its responses and truthful interaction dynamics exist, participants expressed optimism about its future utility.

For future work, we plan to investigate ways to build a design feedback system based on the lessons learned. This research direction will involve enhancing chart comprehension and enriching the knowledge base of LLMs with well-founded rationales, including visualization examples and empirical evidence,  developing user interfaces and interaction designs tailored to the designer's workflow, and evaluating the system's practical applicability.

\begin{acks}
We would like to acknowledge the support of the National Science Foundation (\#2146868).
\end{acks}

\bibliographystyle{ACM-Reference-Format}
\bibliography{refernece}

\appendix
\added{\section{Scoring rubrics and guidelines for LLM-based evaluation}}
\added{In the LLM-based automatic evaluation (Section \ref{sec:llm-based-eval}), we integrate the scoring rubrics and guidelines listed below as a part of the evaluation prompt. While the scoring rubrics were originally defined in Table \ref{tab:rubrics}, we make them more comprehensive and descriptive based on experiments in order to better guide the LLM evaluator.}

\begin{longtable}{@{}p{0.15\textwidth} p{0.83\textwidth}@{}}
\caption{Scoring rubrics for LLM-based evaluation  \label{tab:eval-rubrics}}\\
\toprule
\textbf{Metric} & \textbf{Score \& Description} \\
\midrule
\endfirsthead
\toprule
\textbf{Metric} & \textbf{Score \& Description} \\
\midrule
\endhead
\midrule
\multicolumn{2}{r}{\textit{Continued on the next page}} \\ 
\midrule
\endfoot
\bottomrule
\endlastfoot
\multirow{5}{=}{\textbf{Coverage}}
& \textbf{1:} Addresses 0--20\% of the user’s questions. Almost fails to cover the required points and the majority of the core components, rendering the response ineffective. \\
& \textbf{2:} Addresses 20--40\% of the user’s questions. Misses multiple major components of the question, leaving significant portions of the query unanswered.\\
& \textbf{3:} Addresses 40--60\% of the user’s questions. Covers the majority of aspects but omits several important components, leaving notable gaps in the response coverage. \\
& \textbf{4:} Addresses 60--80\% of the user’s questions. Covers all critical aspects with only minor elements missing, which does not affect the overall completeness of the response. \\
& \textbf{5:} Addresses 80--100\% of the user’s questions. Meticulously addresses every aspect of the question with exhaustive detail, ensuring no possible improvement but with no redundant and lengthy description. \\
\midrule
\multirow{5}{=}{\textbf{Topicality}}
& \textbf{1:} Mostly or entirely off-topic. Fails to maintain any meaningful connection to the subject matter. Major deviation from the question, with significant irrelevant content or a complete failure to address the core question. \\
& \textbf{2:} Partially addresses the topic but frequently digresses. Significant portions stray from the main subject. Noticeable irrelevant points or omissions of several key aspects of the question, leading to a lack of focus. \\
& \textbf{3:} Somewhat relevant; minor deviations from the topic. Contains slight irrelevant content or misses a minor part of the question, but the main focus is generally maintained. \\
& \textbf{4:} Mostly on-topic; occasional tangents. Primarily addresses the question with only a few minor deviations or tangential points, which does not detract from the overall focus. \\
& \textbf{5:} Fully on-topic with no noticeable digressions. Addresses the question with no deviation, and zero irrelevant content. Every sentence directly contributes to answering the question without too much detail. \\
\midrule
\multirow{5}{=}{\textbf{Breadth}}
& \textbf{1:} Very narrow; offering no alternative ideas. Extremely narrow focus missing most relevant aspects. Shows no awareness of alternative perspectives. \\
& \textbf{2:} Covers a small range of ideas, omitting key concepts. Develops these limited ideas superficially without exploring their connections or implications. \\
& \textbf{3:} Moderately diverse but limited in scope. Addresses some relevant aspects but misses several important areas. Shows restricted perspective with minimal variety. \\
& \textbf{4:} Mostly on-topic; occasional tangents. Primarily addresses the question with only a few minor deviations or tangential points, which does not detract from the overall focus. \\
& \textbf{5:} Fully on-topic with no noticeable digressions. Addresses the question with no deviation, and zero irrelevant content. Every sentence directly contributes to answering the question without too much detail. \\ \\
\midrule
\multirow{5}{=}{
  \textbf{Depth}
}
& \textbf{1:} Very minimal detail; lacks substance or justification. Offers completely superficial content. Shows no understanding of core concepts or principles. \\
& \textbf{2:} Superficial treatment with limited insight. Provides superficial explanations missing key concepts. Lacks detailed analysis of the concepts presented. \\
& \textbf{3:} Moderately detailed; some analysis or explanation. Presents basic explanations with limited exploration. Provides surface-level analysis without diving into underlying complexities. \\
& \textbf{4:} Good explanation with solid reasoning and insight. Offers solid explanations of core concepts with meaningful connections. Shows clear understanding beyond surface level. \\
& \textbf{5:} Deeper explanation, demonstrated with examples or expertise. Makes sophisticated connections between concepts and demonstrates expert-level response. \\ \\
\midrule
\multirow{5}{=}{
  \textbf{Clarity}
}
& \textbf{1:} Very unclear; difficult to follow or disorganized. The writing is significantly confusing and often incomprehensible. \\
& \textbf{2:} Somewhat unclear; main ideas partially obscured. The response is poorly organized, with significant logical gaps and a lack of clear structure. Examples, if present, are confusing or irrelevant. \\
& \textbf{3:} Moderately clear; understandable with some effort. The response has noticeable clarity issues, with ideas not flowing smoothly from one to another. Some points require re-reading to grasp their meaning. \\
& \textbf{4:} Clear, well-structured, and easy to read. The response flows smoothly with only occasional minor unclear points. The writing is precise and unambiguous, requiring minimal effort to understand. \\
& \textbf{5:} Flows seamlessly with minimal effort to comprehend. The response is meticulously organized with distinct, logical section breaks. Each section contains precise, directly relevant examples that unequivocally support the points made. \\ \\
\midrule
\multirow{6}{=}{
 \textbf{Actionability}
}
& \textbf{1:} No actionable guidance or next steps. Completely lacks practical implementation value. Offers no useful guidance or implementable steps. Completely impractical advice. \\
& \textbf{2:} Little actionable guidance; mostly explanations. Provides vague guidance without clear steps. Advice is difficult to implement. \\
& \textbf{3:} Some actionable guidance; partially implementable. Presents some practical advice but with unclear steps. Guidance lacks specificity. \\
& \textbf{4:} Clear, actionable guidance that can be followed easily. Offers practical advice with clear steps, though some details may need clarification. \\
& \textbf{5:} Highly actionable; provides explicit step-by-step instructions. Provides specific, implementable steps with practical examples. Anticipates challenges and offers solutions. \\
\end{longtable}

\vspace{2em}
\begin{longtable}{@{}p{0.15\textwidth} p{0.83\textwidth}@{}}
\caption{Scoring guideline for LLM-based evaluation  \label{tab:eval-guidelines}}\\

\toprule
\textbf{Metric} & \textbf{Guideline} \\
\midrule
\endfirsthead
\toprule
\textbf{Metric} & \textbf{Guideline} \\
\midrule
\endhead
\midrule
\multicolumn{2}{r}{\textit{Continued on the next page}} \\ 
\midrule
\endfoot
\bottomrule
\endlastfoot
\multirow{5}{=}{\textbf{Coverage}}
& - Check if all parts of the question are addressed \\
& - Assess completeness of the response relative to the query \\
& - Look for gaps in addressing key aspects of the question \\
& - Evaluate whether important sub-topics are covered \\
& - Check if the scope of the response matches the question \\
\midrule
\multirow{5}{=}{\textbf{Topicality}}
& - Check if the response directly answers the main question \\
& - Identify any off-topic or tangential information \\
& - Assess whether key aspects of the question are addressed \\
& - Look for focus maintenance throughout the response \\
& - Evaluate relevance of any additional context provided \\
\midrule
\multirow{5}{=}{\textbf{Breadth}}
& - Check coverage of different relevant aspects or perspectives \\
& - Assess variety of examples or use cases provided \\
& - Look for consideration of alternative approaches \\
& - Evaluate inclusion of related concepts or implications \\
& - Check if multiple stakeholder viewpoints are considered \\
\midrule
\multirow{5}{=}{
  \textbf{Depth}
}
& - Assess how thoroughly core concepts are explained \\
& - Look for detailed explanations of underlying principles \\
& - Check for discussion of relationships between concepts \\
& - Evaluate the presence of advanced or nuanced insights \\
& - Consider whether complex aspects are adequately explored \\
\midrule
\multirow{5}{=}{
  \textbf{Clarity}
}
& - Evaluate the logical flow and structure \\
& - Check for clear transitions between ideas \\
& - Assess the use of technical terms and their explanations \\
& - Look for concrete examples that illustrate points \\
& - Check if the response is well-organized with clear paragraphs \\
\midrule
\multirow{6}{=}{
  \textbf{Actionability}
}
& - Assess whether advice can be practically implemented \\
& - Look for specific, concrete steps or recommendations \\
& - Check if prerequisites or requirements are clearly stated \\
& - Evaluate the practicality of suggested solutions \\
& - Consider whether guidance is appropriate for the context \\
\end{longtable}

\end{document}